\documentstyle[12pt,graphicx,subfigure]{article}                
\begin{document}
\def\today{\ifcase\month\or
 January\or February\or March\or April\or May\or June\or
 July\or August\or September\or October\or November\or December\fi
 \space\number\day, \number\year}
\def\thebibliography#1{\section*{References\markboth
 {References}{References}}\list
 {[\arabic{enumi}]}{\settowidth\labelwidth{[#1]}
 \leftmargin\labelwidth
 \advance\leftmargin\labelsep
 \usecounter{enumi}}
 \def\newblock{\hskip .11em plus .33em minus .07em}
 \sloppy
 \sfcode`\.=1000\relax}
\let\endthebibliography=\endlist
%
\def\Journal#1#2#3#4{{#1} {\bf #2}, #3 (#4)}

\def\NCA{\em Nuovo Cimento}
\def\NIM{\em Nucl. Instrum. Methods}
\def\NIMA{{\em Nucl. Instrum. Methods} A}
\def\NPA{{\em Nucl. Phys.} A}
\def\NPB{{\em Nucl. Phys.} B}
\def\PLB{{\em Phys. Lett.}  B}
\def\PRL{\em Phys. Rev. Lett.}
\def\PRD{{\em Phys. Rev.} D}
\def\ZPC{{\em Z. Phys.} C}
\def\mgravi{m_{3/2}}
\def\lsim{\ ^<\llap{$_\sim$}\ }
\def\gsim{\ ^>\llap{$_\sim$}\ }
\def\r2{\sqrt 2}
\def\beq{\begin{equation}}
\def\eeq{\end{equation}}
\def\beqn{\begin{eqnarray}}
\def\eeqn{\end{eqnarray}}
\def\rmuu{\gamma^{\mu}}
\def\rmud{\gamma_{\mu}}
\def\PL{{1-\gamma_5\over 2}}
\def\PR{{1+\gamma_5\over 2}}
\def\sinW2{\sin^2\theta_W}
\def\AEM{\alpha_{EM}}
\def\mul{M_{\tilde{u} L}^2}
\def\mur{M_{\tilde{u} R}^2}
\def\mdl{M_{\tilde{d} L}^2}
\def\mdr{M_{\tilde{d} R}^2}
\def\mz2{M_{z}^2}
\def\c2b{\cos 2\beta}
\def\au{A_u}
\def\ad{A_d}
\def\cob{\cot \beta}
\def\v#1{v_#1}
\def\tb{\tan\beta}
\def\epem{$e^+e^-$}
\def\KK{$K^0$-$\bar{K^0}$}
\def\wi{\omega_i}
\def\xj{\chi_j}
\def\Wmu{W_\mu}
\def\Wnu{W_\nu}
\def\m#1{{\tilde m}_#1}
\def\mH{m_H}
\def\mw#1{{\tilde m}_{\omega #1}}
\def\mx#1{{\tilde m}_{\chi^{0}_#1}}
\def\mc#1{{\tilde m}_{\chi^{+}_#1}}
\def\mwi{{\tilde m}_{\omega i}}
\def\mxi{{\tilde m}_{\chi^{0}_i}}
\def\mci{{\tilde m}_{\chi^{+}_i}}
\def\mz{M_z}
\def\sw{\sin\theta_W}
\def\cw{\cos\theta_W}
\def\cb{\cos\beta}
\def\sb{\sin\beta}
\def\rwi{r_{\omega i}}
\def\rxj{r_{\chi j}}
\def\rfp{r_f'}
\def\Kik{K_{ik}}
\def\Fq2{F_{2}(q^2)}
\def\mg{m_{\frac{1}{2}}}
\def\mchi1{m_{\chi}}
\def\tw{\tan\theta_W}
\def\sec2w{sec^2\theta_W}

\begin{center}{\Large \bf Modular Invariant Soft Breaking,
WMAP, Dark Matter and Sparticle Mass Limits}\\
\vskip.25in
{Utpal Chattopadhyay\footnote{E-mail: tpuc@iacs.res.in}$^{(a,b)}$
 and  Pran Nath\footnote{E-mail: nath@neu.edu}$^{(c)}$  }\\
 
{\it
(a) Department of Theoretical Physics, Indian Association for the Cultivation 
of Science, Jadavpur, Kolkata 700032, India\footnote{Permanent address}\\
(b) Theoretical Physics Division, CERN CH 1211, Geneva, Switzerland\\
(c) Department of Physics, Northeastern University, Boston, MA 02115-5005, USA\\
}
\end{center}

\begin{abstract}
An analysis of soft breaking  under the constraint of modular invariance 
is given.  The role 
of dilaton and moduli dependent front factors in achieving a modular 
invariant
$V_{soft}$ is emphasized. Further, it is shown that in string models
$\tan\beta$  is no longer a free parameter but is determined 
in terms of $\alpha_{string}$ and the other soft parameters 
by the constraints of modular invariance and radiative electroweak
symmetry breaking.  The above framework is then used to
analyze the neutralino relic density consistent with the  
WMAP data at self dual points in the K\"ahler and complex structure moduli. 
One finds that the combined set of constraints arising
from modular invariant soft breaking, radiative electroweak 
symmetry breaking and WMAP lead to upper limits 
on sparticle masses for $\mu>0$. These limits are investigated for a 
class of models and found to lie within reach of the Tevatron 
  and of the Large Hadron Collider (LHC). Further, an  
   analysis of the  neutralino-proton cross section shows that 
dark matter in these models can be  explored
in CDMS (Soudan), GENIUS and ZEPLIN. While the analysis is carried
out within the general framework of heterotic strings, 
the possibility that the results of the analysis may extend to a 
broader class  of string models, such as models based on intersecting 
D branes, is also discussed.
 \end{abstract}
 \section{Introduction}
Recently the Wilkinson Microwave Anisotropy Probe 
(WMAP)\cite{bennett,spergel} has measured
the cosmological parameters to a high degree of accuracy leading to
 a very precise constraint on the density of cold dark matter (CDM) 
 in the
 universe. Thus the data suggests an $\Omega_{CDM} h^2$ in the  range
 \beq
\Omega_{CDM}h^2 = 0.1126^{+0.016}_{-0.018}
\label{wmapeqn}
\eeq
 Here $\Omega_{CDM}=\rho_{CDM}/\rho_c$ where  $\rho_{CDM}$ is the mass
 density of the CDM and  $\rho_c$ is the critical mass density needed
 to close the universe, and $h$ is the Hubble parameter in units
 of 100Km/s.Mpc. The implications of this constraint were 
 analyzed in Ref.\cite{Chattopadhyay:2003xi} in the framework 
 of mSUGRA and quite 
 surprisingly it was found that relic density constraint of 
 Eq.~(1) allows for scalar masses to be in the domain
 of several TeV which may even lie beyond the reach of the 
 LHC (For other analyses of WMAP data see 
 Ref.\cite{elliswmap,hb/fp,Chattopadhyay:2003qh}).
 This is so because radiative breaking of the electroweak
 symmetry occurs  in part on the so called 
 Hyperbolic Branch/Focus Point region
 (HB/FP)\cite{Chan:1997bi} 
 which contains an inversion region\cite{Chattopadhyay:2003xi} 
 where typically the Higgs mixing 
 parameter $\mu$ is relatively small, i.e., $|\mu|<< m_0, m_{1/2}$
 where $m_0$ is the universal scalar mass, $m_{1/2}$
 is the universal gaugino mass at the GUT scale $M_G \sim 2 \times 10^{16}$ 
GeV, and $\mu$ is the Higgs mixing parameter which appears in the 
superpotential in the form $\mu H_1H_2$. In the inversion
 region the lightest supersymmetric particle (LSP) is the neutralino
 with mass $\sim |\mu|$ and is  degenerate with the next to the
 lightest neutralino and the light chargino. Consequently there
 is a lot of coannihilation in this region allowing for the satisfaction
 of the WMAP relic density constraints even for very large values of
 $m_0, m_{1/2}$.
 
  In this paper we show that the situation
 is drastically different in a class of string models because of the
 more constrained parameter space\cite{nathtaylor}. 
 Additionally the constraint of
 radiative breaking of the electroweak symmetry is more stringent
 allowing for a determination of $\tan\beta$. 
Our underlying procedure is to consider effective low energy
theory below the string scale for a heterotic 
string model\cite{heterotic}. 
We assume this theory to possess
a T-duality invariance, specifically an SL(2,Z) modular invariance
associated with large-small radius symmetry. Thus we assume that
the effective scalar potential in four dimensions depends on the
dilaton field $S$ and on the K\"ahler moduli fields $T_i$ (i=1,2,3)
and later we will also extend our analysis by including the dependence on the
complex structure moduli $U_i$. 
We require the scalar potential of
the theory to be invariant under the modular $SL(2,Z)$ T transformations
 given by 
 \beqn
  T_i\rightarrow
T'_i=\frac{{a_iT_i-ib_i}}{{ic_iT_i+d_i}},\nonumber\\
\bar T_i\rightarrow
\bar T'_i=\frac{a_i\bar T_i+ib_i}{-ic_i\bar T_i+d_i},\nonumber\\
(a_id_i-b_ic_i)=1,~~~ (a_i,b_i,c_i,d_i \in Z).
\label{mod1}
\eeqn
We follow the approach developed in Ref.\cite{nathtaylor}
(For previous analyses on soft susy breaking in string
 theory  under the constraints of modular invariance see 
 Refs.\cite{fmtv,brignole,nilles,gaillard} and for its phenomenology
 see Refs.\cite{kane}). 
Thus following the supergravity approach adopted in Ref.\cite{can} 
we assume the superpotential to
be composed of a visible sector and a hidden sector so that
$W=W_v+ W_h$ where supersymmetry breaks in the hidden sector 
and is communicated to the visible sector where it leads to generation
of soft breaking.
The  K\"ahler potential is in general given by 
\beqn d=   -ln(S+\bar S +\frac{1}{4\pi^2} \sum_i^3 \delta_i^{GS}
 log(T_i+\bar T_i))
  -\sum_i
log(T_i+\bar T_i) + K_{IJ} Q_I^{\dagger}Q_J +H_{IJ} Q_IQ_J
 \label{kahlerpot}
 \eeqn
where $\delta_i^{GS}$ is the  one loop Green-Schwarz 
correction to the K\"ahler potential\cite{antaylor} and
$Q$ are the matter fields consisting of leptons, quarks and the
Higgs. For the analysis in this work we assume $\delta_i^{GS} =0$. 
The resulting moduli potential have a form which arises in a class
of no scale supergravity [ for a review see \cite{lahanas}]. 
 For the visible sector we assume a general form
$ W_v= \tilde \mu_{IJ} Q_IQ_J + \lambda_{IJK}Q_IQ_JQ_K$.
The outline of the rest of the paper is as follows: In Sec.2 we
give a pedagogical description of modular properties of various
quantities that appear in soft breaking. 
We emphasize the role of dilaton and moduli dependent front factors
in achieving a modular invariant soft breaking and an explicit proof 
of modular invariance of $V_{soft}$ is also given.  
In Sec.3 we discuss radiative breaking of the electroweak symmetry
in the context of string models and show that $\tan\beta$ is 
not a free parameter but a determined quantity in string models.
In Sec.4 we use the modular invariant soft breaking and the determination of
$\tan\beta$ to compute the neutralino relic density in the range
allowed by the WMAP data. 
The cases $\mu>0$ and $\mu<0$ 
are found to have significantly different behaviors.
For the case $\mu>0$ (which appears to be the preferred value 
of $\mu$\cite{Chattopadhyay:2002jx} from the 
$g_{\mu}-2$ data\cite{Bennett:2004pv})
it  is shown that the modular
invariant soft breaking, the string determined value of $\tan\beta$ 
 and the WMAP constraint combine to
produce upper limits on sparticle masses which appear to lie 
within reach of the hadron colliders for a class of string
models. Implications for the direct detection of dark matter are
also discussed. Conclusions are given  in Sec.5. 

\section{Modular Invariant Soft Breaking}
In this section we give a pedagogic discussion of the modular
properties of the soft parameters and also show explicitly that
the soft breaking scaler potential is modular invariant.
We begin by defining modular weights.
Suppose we have a function $f(T_i,\bar T_i)$ which transforms 
under modular transformations of  Eq.~(\ref{mod1}) so that 
\beqn
f(T_i,\bar T_i) \rightarrow f(T_i',\bar T_i') 
=(icT_i+d)^{n_1} (-ic\bar T_i+d)^{n_2} f(T_i,\bar T_i) 
\label{mod2}
\eeqn
then it has modular weight $(n_1,n_2)$.
For example,  from  Eq.~(\ref{mod1}) one finds that  
$(T_i+\bar T_i)$$ \rightarrow$$ (T_i'+\bar T_i')$$
= (ic_iT_i+d_i)^{-1}(-ic_i\bar T_i+d_i)^{-1}(T_i+\bar T_i)$
and thus $(T_i+\bar T_i)$ has modular weights $(-1,-1)$.
Now $G$ defined by $G=d+ln(WW^{\dagger})$ is invariant under
modular transformations which implies that $W$ has modular weight
$(-1,0)$ and $W^{\dagger}$ has modular weight $(0,-1)$ under the
modular transformation of Eq.~(\ref{mod1}). Thus 
$WW^{\dagger}$ has modular weight $(-1,-1)$ and
$|W|$ has modular weight (-1/2,-1/2). Defining $\theta_W$ so that
\beqn
W=|W|e^{i\theta_W} 
\eeqn
 we find that $e^{i\theta_W}$ has modular weight (-1/2, 1/2). 
 The Dedekind function $\eta(T_i)$ 
 (where $\eta (T_i)=q^{1/24}\Pi_n(1-q^n)$, $q=e^{2\pi i T_i}$)  
 has modular weight (1/2,0)
 while $G_2(T_i)$ defined by 
 \beqn
 G_2(T_i)= \frac{\partial ln\eta^2(T_i)}{\partial T_i} +\frac{1}{T_i+\bar T_i} 
 \eeqn
has modular weight (2,0). More generally if we have a function 
$Q(T_i)$ with modular weight $(n_Q^i, 0)$ then the covariant derivative
defined by 
\beqn
D_{T_i}[ln Q(T_i)] \equiv \frac{\partial ln Q(T_i)}{\partial T_i} +
\frac{n_Q^i}{T_i+\bar T_i} 
\eeqn
transforms as 
\beqn
D_{T_i}[ln Q(T_i)] \rightarrow D_{T_i'}[ln Q(T_i')]
=(icT_i+d)^2 D_{T_i}[ln Q(T_i)]  
\eeqn
and thus $D_{T_i}[ln Q(T_i)]$ has modular weight $(2,0)$.

In Table 1 we summarize modular weights of several quantities
that appear in the analysis of soft breaking.
We consider now the specific model of  
Ref.\cite{nathtaylor}. For $W_h$ we assume that the modular 
weight is carried by
the Dedekind function $\eta (T_i)$  so that 
$W_h={F(S,T_i)}/{\prod_{i=1}^3\eta(T_i)^{2}}$,  
where $F(S,T_i)$ is modular invariant. 
For the  K\"ahler potential we consider the specific form
$d=D(S,\bar S)$ -$\sum_iln(T_i+\bar T_i) $+$
\sum_{i\alpha}(T_i+\bar T_i)^{n_{\alpha}^i}$ 
$C_{\alpha}^{\dagger}C_{\alpha}$, 
where $C_{\alpha}$ are the charged fields with modular weights 
$n^i_{\alpha}$. 
We introduce the  notation $\gamma_s= (S+\bar S)G,S/\sqrt 3$ 
$= |\gamma_S| e^{i\theta_S}$,
 $\gamma_{T_i}= (T_i+\bar T_i)G,{T_i}/\sqrt 3$ 
 $= |\gamma_{T_i}| e^{i\theta_{T_i}}$. The condition that
 vacuum energy vanish then takes the form
 \beqn
 |\gamma_S|^2+\sum_{i=1}^3 |\gamma_{T_i}|^2 =1
 \label{vaccond}
 \eeqn

\begin{table}[hbt]
{\centering 
\begin{tabular}{|c|c|}
\hline
quantity   &  modular weights \\
\hline
$|W|$                          &  $(-\frac{1}{2}, -\frac{1}{2})$    \\
\hline
$e^{i\theta_W}$     &   $(-\frac{1}{2}, \frac{1}{2})$       \\
\hline
 $\eta(T_i)$   &      $(\frac{1}{2}, 0)$                \\
\hline
 $2\partial_{T_i}ln\eta(T_i)+(T_i+\bar T_i)^{-1}$  & $(2,0)$    \\
\hline
  $\partial_{T_i} W - (T_i+\bar T_i)^{-1}W$    & $(1,0)$   \\
\hline
  $Y_{\alpha\beta\gamma}(T_i)$ & $(-1-n_{\alpha\beta\gamma}^i, 0)$ \\
 \hline
  $\lambda_{\alpha\beta\gamma}(T_i)$ & $(0,0)$ \\ 
\hline
  $\partial_{T_i}lnY_{\alpha\beta\gamma}(T_i)- (1+n_{\alpha\beta\gamma}^i)
  (T_i+\bar T_i)^{-1}$  &  $(2,0)$    \\
 \hline
  $\mu_{\alpha\beta}(T_i)$ & $(-1-n_{\alpha\beta}^i, 0)$                     \\
\hline 
$\tilde\mu_{\alpha\beta}(T_i)$ & $(0,0)$ \\
\hline
  $\partial_{T_i}ln\mu_{\alpha\beta}(T_i)- (1+n_{\alpha\beta}^i)
  (T_i+\bar T_i)^{-1}$  &  $(2,0)$    \\
  \hline
$\partial_{T_i}\lambda_{\alpha\beta\gamma}(T_i)$ & $(2,0)$ \\  
\hline
$\partial_{T_i}\tilde\mu_{\alpha\beta}(T_i)$ & $(2,0)$ \\  
  \hline
 $(T_i+\bar T_i)$ & $(-1,-1)$ \\
 \hline
  $|\gamma_s|$  & $(0,0)$ \\
\hline
  $|\gamma_{T_i}|$       &  $(0,0)$                   \\
\hline
   $e^{i\theta_{T_i}}$   &   $(1,-1)$    \\
\hline
 $e^{i\theta_{S}}$   &   $(0,0)$    \\
\hline
$A^{0}_{\alpha\beta\gamma}$ & $(1,0)$ \\
\hline
$B^{0}_{\alpha\beta}$ & $(1,0)$ \\
\hline
$1/\sqrt{f}$  &  ($\frac{1}{2}$,$\frac{1}{2}$) \\
\hline                      
\hline
\end{tabular}
\par}
\centering
\caption{A list of modular weights under the modular 
transformations of Eq.~(\ref{mod1})}
\end{table}

The soft breaking potential for this case  is given by 
\beq
V_{soft}=m_{3/2}^2 \sum_{\alpha}(1+3\sum_{i=1}^3n^i_{\alpha}|\gamma_{T_i}|^2)
c_{\alpha}^{\dagger}c_{\alpha}+ (\sum_{\alpha\beta}B^0_{\alpha\beta}w^{(2)}_{\alpha\beta} + 
\sum_{\alpha\beta\gamma} 
A_{\alpha\beta\gamma}^0w_{\alpha\beta\gamma}^{(3)}+ H.c.)
\eeq
where the normalized fields $c_{\alpha}$ and $m_{3/2}$ ($=e^{-G/2}$)  
are modular invariant and $w^{(2)}_{\alpha\beta} = \mu_{\alpha\beta}
C_{\alpha}C_{\beta}$  and 
$w^{(3)}_{\alpha\beta\gamma} = Y_{\alpha\beta\gamma}
C_{\alpha}C_{\beta}C_{\gamma}$. 
Following the standard procedure the soft breaking parameters
$A^0$ and $B^0$ may be expressed in the form
\beqn 
A_{\alpha\beta\gamma}^0= -\sqrt 3 m_{3/2} \frac{e^{D/2-i\theta_W}}{\sqrt f}  
[ |\gamma_{S}|  e^{-i \theta_S} 
 (1-(S+\bar S)\partial_S lnY_{\alpha\beta\gamma})\nonumber\\ 
 +\sum_{i=1}^3 |\gamma_{T_i}| 
e^{-i \theta_{T_i}}(1+n^i_{\alpha}+n^i_{\beta} +n^i_{\gamma} 
 -(T_i+\bar T_i)\partial_{T_i}ln Y_{\alpha\beta\gamma}
 -(T_i+\bar T_i) n^i_{\alpha\beta\gamma} G_2(T_i))] 
\label{A1}
\eeqn
and 
\beqn 
B_{\alpha\beta}^0= -m_{3/2} \frac{e^{D/2-i\theta_W}}{\sqrt f}  
[ 1+ \sqrt 3  |\gamma_{S}|  e^{-i \theta_S} 
 (1-(S+\bar S)\partial_S ln\mu_{\alpha\beta})\nonumber\\ 
 + \sqrt 3\sum_{i=1}^3 |\gamma_{T_i}| 
e^{-i \theta_{T_i}}(1+n^i_{\alpha}+n^i_{\beta} 
 -(T_i+\bar T_i)\partial_{T_i}ln\mu_{\alpha\beta}
 -(T_i+\bar T_i) n^i_{\alpha\beta} G_2(T_i)
 )] 
\label{B1}
\eeqn
The modular properties of $A_{\alpha\beta\gamma}^0$ and 
$B_{\alpha\beta}^0$ are not so manifest as written in the above
form. This is so because 
$\partial_{T_i}(ln Y_{\alpha\beta\gamma})$ and 
$\partial_{T_i}(ln \mu_{\alpha\beta})$ have no well defined 
modular weights. Indeed they are mixtures of terms with 
different modular weights.  Thus we may express them as
follows 
\beqn
\partial_{T_i}(ln Y_{\alpha\beta\gamma})= 
D_{T_i}(ln Y_{\alpha\beta\gamma}) 
+(1+n^i_{\alpha\beta\gamma}) (T_i+\bar T_i)^{-1}\nonumber\\
\partial_{T_i}(ln \mu_{\alpha\beta}) =
D_{T_i}(ln \mu_{\alpha\beta}) +(1+n^i_{\alpha\beta}) (T_i+\bar T_i)^{-1}
\nonumber\\
n^i_{\alpha\beta\gamma} =n^i_{\alpha} + n^i_{\beta} + n^i_{\gamma},
~~ n^i_{\alpha\beta} =n^i_{\alpha} + n^i_{\beta} 
\label{identity1}
\eeqn
where $D_{T_i}(ln Y_{\alpha\beta\gamma})$ and 
$D_{T_i}(ln \mu_{\alpha\beta})$ have modular weight $(2,0)$
while $(T_i+\bar T_i)^{-1}$ has  modular weight ($1,1)$.
Thus we see that $\partial_{T_i}(ln Y_{\alpha\beta\gamma})$ and 
$\partial_{T_i}(ln \mu_{\alpha\beta})$ are 
a linear combination of quantities with modular weight 
(2,0) and (1,1). Using Eq.~(\ref{identity1}) we may express 
Eqs.~(\ref{A1}) and  (\ref{B1}) in the form
\beqn 
A_{\alpha\beta\gamma}^0= m_{3/2} \frac{e^{D/2-i\theta_W}}{\sqrt f}  
\sqrt 3 \sum_{i=1}^3  |\gamma_{T_i}| (T_i+\bar T_i)
e^{-i \theta_{T_i}} [n^i_{\alpha\beta\gamma} G_2(T_i) + D_{T_i} 
ln Y_{\alpha\beta\gamma}] \nonumber\\
-m_{3/2} \frac{e^{D/2-i\theta_W}}{\sqrt f} 
\sqrt 3 |\gamma_{S}|  e^{-i \theta_S} 
 (1-(S+\bar S)\partial_S lnY_{\alpha\beta\gamma}) 
\label{A2}
\eeqn
and
\beqn 
B_{\alpha\beta}^0= m_{3/2} \frac{e^{D/2-i\theta_W}}{\sqrt f}  
\{-1+
\sqrt 3 \sum_{i=1}^3 |\gamma_{T_i}| (T_i+\bar T_i)
e^{-i\theta_{T_i}} [n^i_{\alpha\beta} G_2(T_i) + D_{T_i} 
ln\mu_{\alpha\beta}]\} \nonumber\\
-m_{3/2} \frac{e^{D/2-i\theta_W}}{\sqrt f} 
\sqrt 3 |\gamma_{S}|  e^{-i\theta_S} 
 (1-(S+\bar S)\partial_S ln\mu_{\alpha\beta}) 
\label{B2}
\eeqn
The modular properties
of $A^0$ and $B^0$ can be easily read off from above.
Using Table 1, we find that in Eq.~(\ref{A2}) the front factor 
$e^{D/2-i\theta_W}/\sqrt f$ has a modular (1,0) and the 
remaining factors have modular weight (0,0) giving $A^0$
an overall modular weight (1,0) as desired. The same also holds
for $B^0$ as can be read off from Eq.(\ref{B2}). While  
expressions of Eqs.(\ref{A2}) and (\ref{B2}) are  useful in 
exhibiting the modular properties of $A^0$ and $B^0$,
they (as well as 
Eqs.~(\ref{A1}) and (\ref{B1})) seem to imply at least superficially 
that $A^0$ and $B^0$ have a significant dependence on the modular weights.
This is not really the case as we now demonstrate. To this end 
we define modular invariant couplings 
 $\lambda_{\alpha\beta\gamma}$ and $\tilde \mu_{\alpha\beta}$ 
 so that
 \beqn 
 Y_{\alpha\beta\gamma} =\lambda_{\alpha\beta\gamma}
\eta^{-2(1+n^i_{\alpha\beta\gamma})}(T_i), ~~
=\mu_{\alpha\beta} = \tilde \mu_{\alpha\beta}
\eta^{-2(1+n^i_{\alpha\beta})}(T_i)
\eeqn
 where $\lambda_{\alpha\beta\gamma}$ and $\tilde \mu_{\alpha\beta}$
have modular weights $(0,0)$. 
We note the following identity
\beqn
\partial_{T_i} ln Y_{\alpha\beta\gamma} = 
\partial_{T_i} ln \lambda_{\alpha\beta\gamma} 
-(1+n^i_{\alpha} + n^i_{\beta} +n^i_{\gamma})G_2(T_i) + 
\frac{1+ n^i_{\alpha} + n^i_{\beta} +n^i_{\gamma}}{T_i+\bar T_i} 
\label{identity2}
\eeqn
which can be gotten by using  the relation
$\partial_{T_i} ln\lambda_{\alpha\beta\gamma}
=D_{T_i}(ln Y_{\alpha\beta\gamma}) +(1+n^i_{\alpha\beta\gamma})G_2(T_i)$
and Eq.~(\ref{identity1}).
Using Eq.~(\ref{identity2}) in Eq.~(\ref{A1}) we may write 
$A_{\alpha\beta\gamma}^0$  as follows
\beqn 
A_{\alpha\beta\gamma}^0= m_{3/2} \frac{e^{D/2 -i\theta_W}}{\sqrt f}  
\sqrt 3 \sum_{i=1}^3 |\gamma_{T_i}| (T_i+\bar T_i)
e^{-i \theta_{T_i}} [-G_2(T_i) + \partial_{T_i} 
ln\lambda_{\alpha\beta\gamma}] \nonumber\\
-m_{3/2} \frac{e^{D/2-i\theta_W}}{\sqrt f} 
\sqrt 3 |\gamma_{S}|  e^{-i\theta_S} 
 (1-(S+\bar S)\partial_S lnY_{\alpha\beta\gamma}) 
\label{A3}
\eeqn
and a similar analysis gives
\beqn 
B_{\alpha\beta}^0= m_{3/2} \frac{e^{D/2-i\theta_W}}{\sqrt f}  
\{-1+
\sqrt 3 \sum_{i=1}^3|\gamma_{T_i}| (T_i+\bar T_i)
e^{-i\theta_{T_i}} [-G_2(T_i) + \partial_{T_i} 
ln\tilde\mu_{\alpha\beta}]\} \nonumber\\
-m_{3/2} \frac{e^{D/2-i\theta_W}}{\sqrt f} 
\sqrt 3 |\gamma_{S}|  e^{-i \theta_S} 
 (1-(S+\bar S)\partial_S ln\mu_{\alpha\beta}) 
\label{B3}
\eeqn
 In Eqs.~(\ref{A3}) and ~(\ref{B3})
 one finds that the modular weights have disappeared due to
 the cancellation arising out of the identity Eq.~(\ref{identity2}).
Further,  using Table 1 we find that the modular weight of $A^0$ is
 manifestly (1,0) and similarly the modular  weight of $B^0$ is (1,0)
  as necessary to achieve a modular invariant 
 $V_{soft}$.
 The importance of the front factors $e^{D/2-i\theta_W}/\sqrt f$ was first 
 emphasized in the work of Ref.\cite{nathtaylor}. This factor is often suppressed
 or omitted in string based analyses of soft breaking. 
 However, since it has a non vanishing modular weight,
 the modular invariance of $V_{soft}$ cannot be maintained without it. 
 Further, the front factor $e^{D/2}$  is also significant in another 
 context as will be discussed in Sec.3.

The analysis given above is more general than of Ref.~(\cite{nathtaylor}).
We can limit to that result by making a further dynamical assumption.
Thus while $\lambda_{\alpha\beta\gamma}$ and $\tilde\mu_{\alpha\beta}$
have modular weights $(0,0)$ they could still have modular 
dependence through dependence on modular invariants. A significant 
simplification occurs if we assume that this dependence is identical
to that of the function $F(S,T_i)$. Specifically we assume that
$\partial_{T_i} \lambda_{\alpha\beta\gamma}= \partial_{T_i} F(S,T_i)$ and 
$\partial_{T_i} \tilde\mu_{\alpha\beta}= \partial_{T_i} F(S,T_i)$.
Under this assumption and also assuming that the potential is minimized at 
the self dual points  $T_i=(1, e^{i\pi/6})$ one finds\cite{nathtaylor} 
\beqn 
A_{\alpha\beta\gamma}^0= m_{3/2} e^{D/2-i\theta_W}
 (3\sum_{i=1}^3 |\gamma_{T_i}|^2 -\sqrt 3 |\gamma_S|
 (1-(S+\bar S)\partial_S lnY_{\alpha\beta\gamma}) 
  e^{-i\theta_s})/f_{\alpha}^{\frac{1}{2}}
\label{Aeqn}
\eeqn
\beqn
 B^0_{\alpha\beta}= m_{3/2} e^{D/2-i\theta_W}
 (-1+3\sum_{i=1}^3 |\gamma_{T_i}|^2 -\sqrt 3 |\gamma_S|
 (1-(S+\bar S)\partial_S ln\mu) 
  e^{-i\theta_s})/f_{\alpha}^{\frac{1}{2}}
\label{Beqn}
\eeqn
where  $f_{\alpha}$ arises due to the possibility of several 
degenerate vacua so that $\{f_{\alpha}\}=8,4\sqrt 3$, $6$,
$3\sqrt 3$.
We can easily extend the above analysis to include the complex 
structure moduli $U_i$ (i=1,2,3) which have $SL(2,Z)$ 
modular transformations of their own under which the scalar
potential is invariant. Including the $U_i$ moduli the 
K\"ahler potential is of the form
\beqn d=   -ln(S+\bar S) -\sum_{i=1}^3 log(T_i+\bar T_i)
  -\sum_{i=1}^3 log(U_i+\bar U_i) 
 \label{umoduli}
 \eeqn
It is convenient to relabel the $U_i$ moduli so that
$T_{3+i}=U_i$ (i=1,2,3). Then assuming that the potential is 
minimized again at the self dual points for all $T_i$, 
 Eqs.~(\ref{Aeqn}) and ~(\ref{Beqn}) are valid with i summed from $1-6$
 where $f_{\alpha}$ assume the set of values $2^n3^{3-\frac{n}{2}}$
 (n=0,..,6). Further, in Eq.~(\ref{vaccond}) the sum over i runs 
 from $1-6$. 
Gauginos also acquire masses after spontaneous breaking of 
supersymmetry so that 
$  M_{\alpha}= \frac{1}{2} {Ref_{\alpha}}^{-1}
  e^{-G/2} f_{\alpha a}(G^{-1})^a_b G^b$,
  where  $f_{\alpha}$ is the  gauge kinetic
  energy function for the case of a product gauge group
  $\cal G=\prod \cal G_{\alpha}$ 
  and has the expansion 
  $f_{\alpha} = k_{\alpha} S + \sum\frac{1}{4\pi^2}[C({\cal G}_{\alpha})
  -\sum_I T(R^{\alpha}_{Q_I})(1+2n^i_{Q_I}) -
  2k_{\alpha} \delta^{GS}_i]log(\eta (T_i)) + \dots,$ 
  where $k_{\alpha}$ is the Kac-Moody level for ${\cal 
G}_{\alpha}$. In our analysis we limit ourselves to Kac-Moody
level 1 and to the universal case so that
  \beqn
  m_{1/2}= \sqrt 3 m_{3/2} |\gamma_S| e^{- i\theta_s}
  \eeqn
 \section{Determination of $\tan\beta$ from Modular Invariance 
 and EWSB Constraints}
Radiative electroweak symmetry breaking (EWSB)  constraints are more 
stringent in
string theory than in supergravity. Radiative symmetry breaking 
produces two constraints arising from the minimization with
respect to the vacuum expectation values of the two Higgs fields
$v_1=<H_1^0>$ and $v_2=<H_2^0>$ of the minimal supersymmetric standard model
(MSSM), i.e., $\partial V/\partial v_i^0 =0$. Typically in supergravity
one of these is used to determine $\mu$ and the other one to 
eliminate the parameter  $B_0$ in terms of $\tan\beta$. 
Now in string theory one has in principle a determination of the
parameter $\mu$ and the fact that it 
should equal the $\mu$ determined
via radiative breaking of the electroweak symmetry would be highly
constraining. In practice there is no realistic string determination of
 $\mu$  (see, however, Ref.\cite{Antoniadis:1994hg}),
 and thus pending such a determination we will
continue to use  the radiative symmetry breaking equation for
the  computation of the $\mu$  parameter. The second symmetry breaking
constraint which eliminates $B_0$ in favor of $\tan\beta$ (see, e.g., 
Ref.\cite{an}) as a 
parameter in supergravity must be treated differently in string theory
since one has a determination of $B_0$ here. A remarkable aspect of
the $B_0$ determination is that it has a front factor of $e^{D/2}$
which is related to the string constant via the relation  
$e^{-D}=\frac{2}{g_{string}^2}$ where $g_{string}= k_ig_i$ and 
$k_i$ is the Kac-Moody level of the
gauge group ${\cal G}_i$ and $g_i$ is the corresponding gauge coupling
constant. Thus the second radiative symmetry breaking equation can
be thought of as a determination of the string constant in the terms
of parameters at the electroweak scale. Alternately we  can think
of this second equation as a determination of $\tan\beta$ so that
\beqn
\tan\beta=
\frac{(\mu^2+ \frac{1}{2} M_Z^2 +m_{H_1}^2) f_{\alpha}^{1/2} } 
{\sqrt{2\pi}\mu m_{3/2} \tilde r_B\alpha_{string} }
(|-1+3\sum_i |\gamma_i|^2-\sqrt 3|\gamma_S|
(1-(S+\bar S)\partial_S ln\mu)|)^{-1}
\label{tan}
\eeqn 
where $\tilde r_B=r_B(1+\epsilon_Z)$ is the renormalization group 
coefficient that relates $B$ at the electroweak scale to $B_0$ at 
the unification  scale
so that  $B=r_B B_0$ and 
 $\epsilon_Z= M_Z^2/(m_{H_1}^2 +\mu^2 -\frac{1}{2}M_Z^2
+(m_{H_1}^2 +\mu^2 +\frac{1}{2}M_Z^2)\cot^2\beta)$.
In this analysis we adopt the procedure of using 
Eq.~(\ref{tan}) to determine $\tan\beta$.

However, care should be taken in implementing Eq.(\ref{tan}).
Since Eq.(\ref{tan}) arises from minimization of   
$V_{Higgs}$ at a higher scale $Q$, i.e., $Q\sim m_{\tilde t}$ or 
$Q\sim$ (highest mass of the spectrum)/2~\cite{Chan:1997bi}, 
where the correction to 
$\mu^2$ from the one loop 
correction of the effective potential is small, 
one has to consider the value of 
 $m_{H_1}^2$,  $m_{H_2}^2$, 
and $\mu^2$  at this scale $Q$ while using Eq.(\ref{tan}).
 This is important since $m_{H_1}^2$ and $m_{H_2}^2$ can vary a lot 
 between $Q\sim M_Z$ and $Q\sim  m_{\tilde t}$ etc.
In this analysis considering one loop correction to the 
effective potential we have adopted an iterative procedure to 
determine $\tan\beta$. This is done by 
starting with a guess value of $\tan\beta$ then using 
it to compute $B_0$ (via $B(Q)$) from the radiative electroweak 
symmetry breaking equation and then comparing 
with the string determined value of $B_0$. 
Such iterations produce rapid convergence of the $B_0$ value 
as obtained from the radiative electroweak symmetry breaking toward  
its string value. Thus $\tan\beta$ is determined for 
given values of the soft parameters, input moduli etc., i.e., the 
choice of the self dual point and $\alpha_{string}$. 

\section{Analysis of Dark Matter and Sparticle Masses 
   at Self Dual Points with WMAP Constraints}
\begin{figure}                       
\vspace*{-0.6in}                                 
\subfigure[
Contours of constant $\tan\beta$, $A_0$ and 
$\mu$ are shown on the available parameter space 
for $\mu>0$ in the $(\gamma_s-m_{3/2})$ plane. 
The $b \rightarrow s + \gamma$ contour 
is shown as a dot-dashed line below which the region is 
disallowed. WMAP satisfied relic density region is shown as small 
shaded area in black. The gray region-I refers to the  discarded 
zone of very large $\tan\beta$ where 
Yukawa couplings are beyond the perturbative domain.
The gray region II results from 
the absence of REWSB or smaller $m_{\tilde \chi_1^\pm}$ below 
the experimental limit.  
]{
\label{gamma_gravi}             
\hspace*{-0.4in}                               
\begin{minipage}[b]{\textwidth}                       
\centering                      
\includegraphics[width=0.8\textwidth, height=0.4\textwidth]{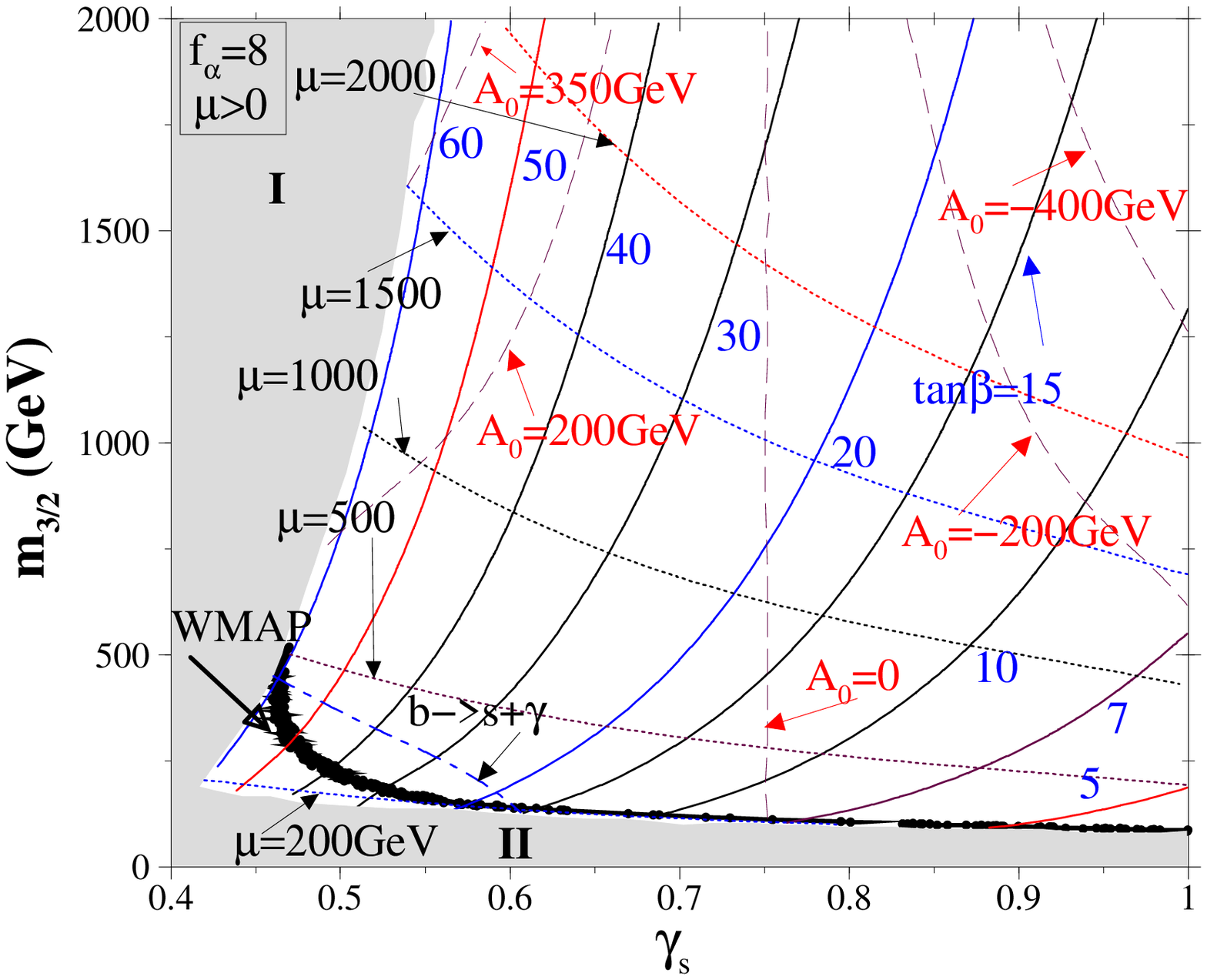} 
\end{minipage}}                       

\subfigure[The contours of constant $m_{3/2}$ in the 
$\gamma_s-\tan\beta$ plane for $\mu>0$ are as labelled.  The 
$b \rightarrow s + \gamma$ disallowed region is in the left of the 
blue dashed line. The WMAP allowed region is shown in narrow shaded area in 
black.]{
\label{gamma_tan}             
\hspace*{-0.4in}                               
\begin{minipage}[b]{\textwidth}                       
\centering                      
\includegraphics[width=0.8\textwidth, height=0.4\textwidth]{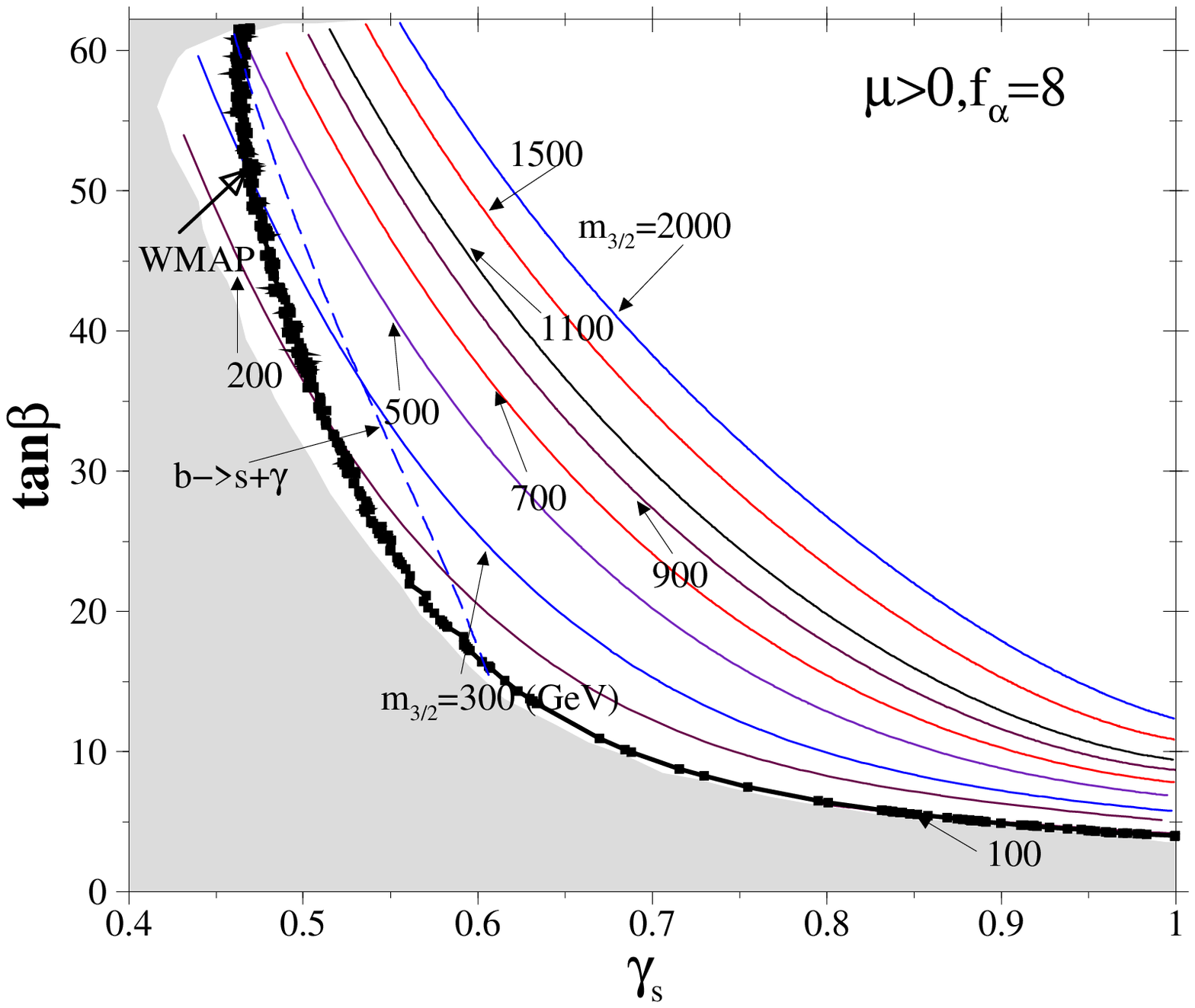} 
\end{minipage}}
\caption{}    
\label{gamma_vs_gravi_tan}                   
\end{figure}

We discuss now the numerical results of the analysis.
For simplicity we will set $n_{\alpha}^i=0$, assume no 
dependence on the moduli $U_i$ (the dependence on the $U_i$ moduli will
be considered later) and further  
assume that one is at the self dual point given by 
$f_{\alpha}=8$. In the analysis we set all the CP phases to zero.
In Fig.~\ref{gamma_gravi}, for $\mu>0$, we give a plot of the self 
consistent determination of $\tan\beta$ and exhibit it as contours
in the $\gamma_s- \mgravi$ plane for values of $\mgravi$
ranging up to 2 TeV.  
One finds a steady decrease of $\tan\beta$ as $\gamma_s$ is increased for 
a given $\mgravi$. 
Contours of $A_0$ and $\mu$
are also shown. 
 Regarding $\mu$, for a given $\mgravi$ (i.e., for a given value of the
universal scalar mass at $M_G$)    
$\mu$ increases with increasing $\gamma_s$ because 
 the  universal gaugino mass at $M_G$ increases  with increasing
 $\gamma_S$. Further, $\mu$ also increases with  
increasing $\mgravi$ for fixed $\gamma_s$ since both the universal 
scalar mass and the universal gaugino mass at $M_G$ 
increase with increasing $\mgravi$.      
In the analysis we also impose the flavor changing neutral current
(FCNC) constraint from the  process 
$b\rightarrow s+\gamma$\cite{bsgamma,bsgammanew,gambino,cleo}
for which we  take the range  $2 \times 10^{-4} < 
Br(b \rightarrow s+ \gamma) < 4.5 \times 10^{-4}$.
In Fig.~\ref{gamma_gravi} the FCNC constraint is shown 
as a dot-dashed line below which the region is 
disallowed.  
Remarkably one finds that the WMAP relic density limits 
($0.094<\Omega_{\tilde \chi_1^0}h^2<0.129 $) 
are satisfied within this 
very constrained system and most of the allowed parameter space is
in the dilaton dominated region with $\gamma_s > 0.6$. 
 The gray region-I in Fig.~\ref{gamma_gravi} refers to the 
the non-perturbative zone where the third generation of 
Yukawa couplings no longer stay within the perturbative 
limits because of very large $\tan\beta$ values. 
The gray region II results from 
the absence of radiative electroweak symmetry breaking 
(REWSB) condition or smaller $m_{\tilde \chi_1^\pm}$ below 
the experimental lower limit.  
In Fig.~\ref{gamma_tan} we exhibit the satisfaction of the WMAP relic density 
constraints in the $\gamma_s-\tan\beta$ plane for $\mu>0$. The contours of 
$\mgravi$ are exhibited and so is the 
FCNC constraint from  $b\rightarrow s+\gamma$  (the disallowed region is 
in the left of the dashed line).  
From Figs.~\ref{gamma_gravi} and \ref{gamma_tan} 
one finds that $\mgravi$ consistent
with WMAP constraint increases with increasing $\tan\beta$.
 However, even with $\tan\beta$ as large as  $60$  one finds that 
$\mgravi$ is no larger than 500 GeV in order that 
one satisfy simultaneously the WMAP constraints and the 
requirement of $\tan\beta$ becoming not too large so that all the 
Yukawa couplings stay within the perturbative domain.

\begin{figure}                       
\vspace*{-0.6in}                                 
\subfigure[Variation of sparticle masses with respect to 
$\gamma_s$ when $m_{3/2}$ is fixed at 500 GeV for $\mu>0$.
The WMAP constraint is not exhibited. ]{
\label{gamma_vs_mass}             
\hspace*{-0.4in}                               
\begin{minipage}[b]{\textwidth}                       
\centering                      
\includegraphics[width=0.9\textwidth, height=0.5\textwidth]{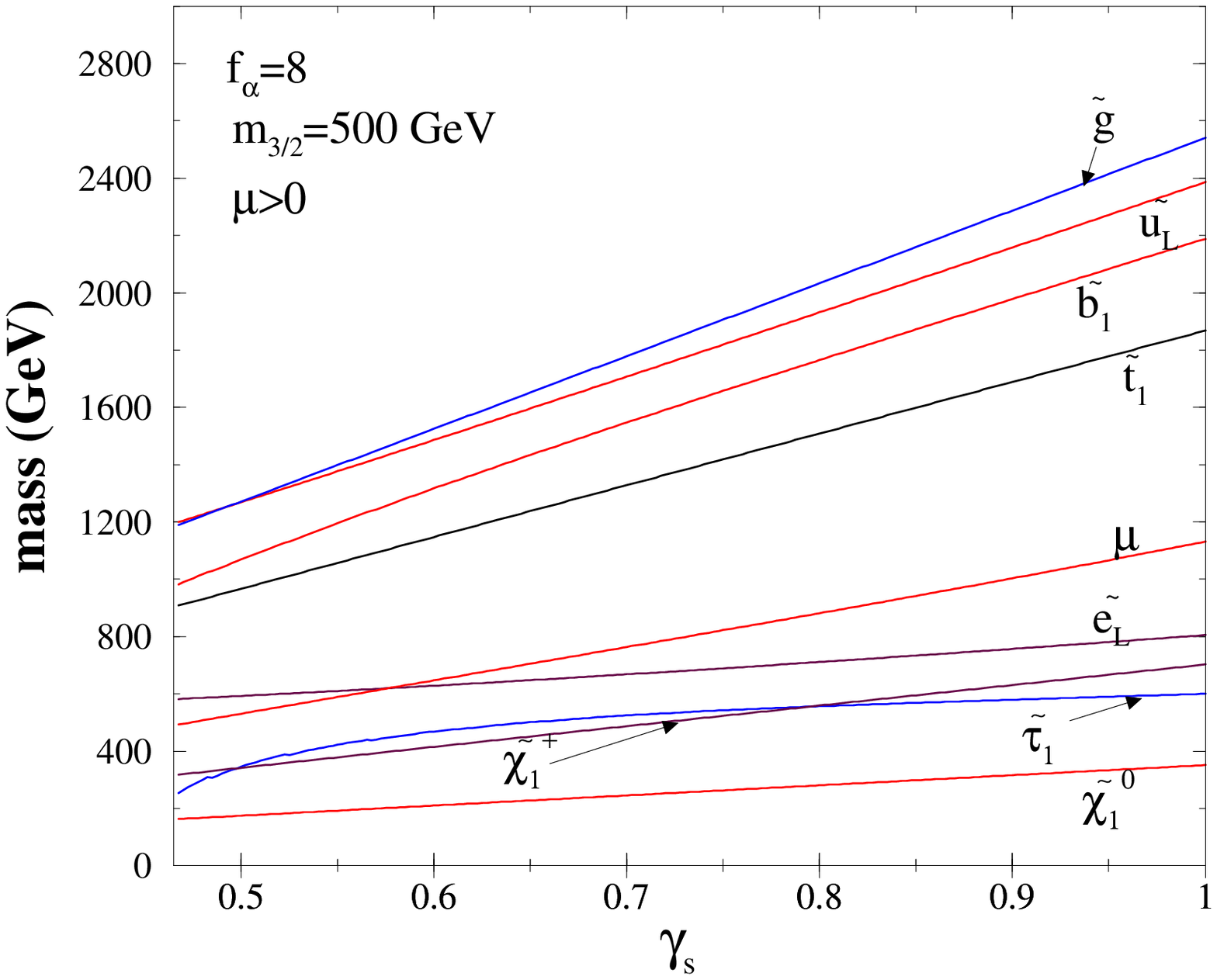} 
\end{minipage}}                       

\subfigure[
Variation of sparticle masses with respect to 
$m_{3/2}$ when $\gamma_s$ is fixed at 0.75 for $\mu>0$.
The WMAP constraint is not exhibited.]{
\label{gravi_vs_mass}             
\hspace*{-0.4in}                               
\begin{minipage}[b]{\textwidth}                       
\centering                      
\includegraphics[width=0.9\textwidth, height=0.5\textwidth]{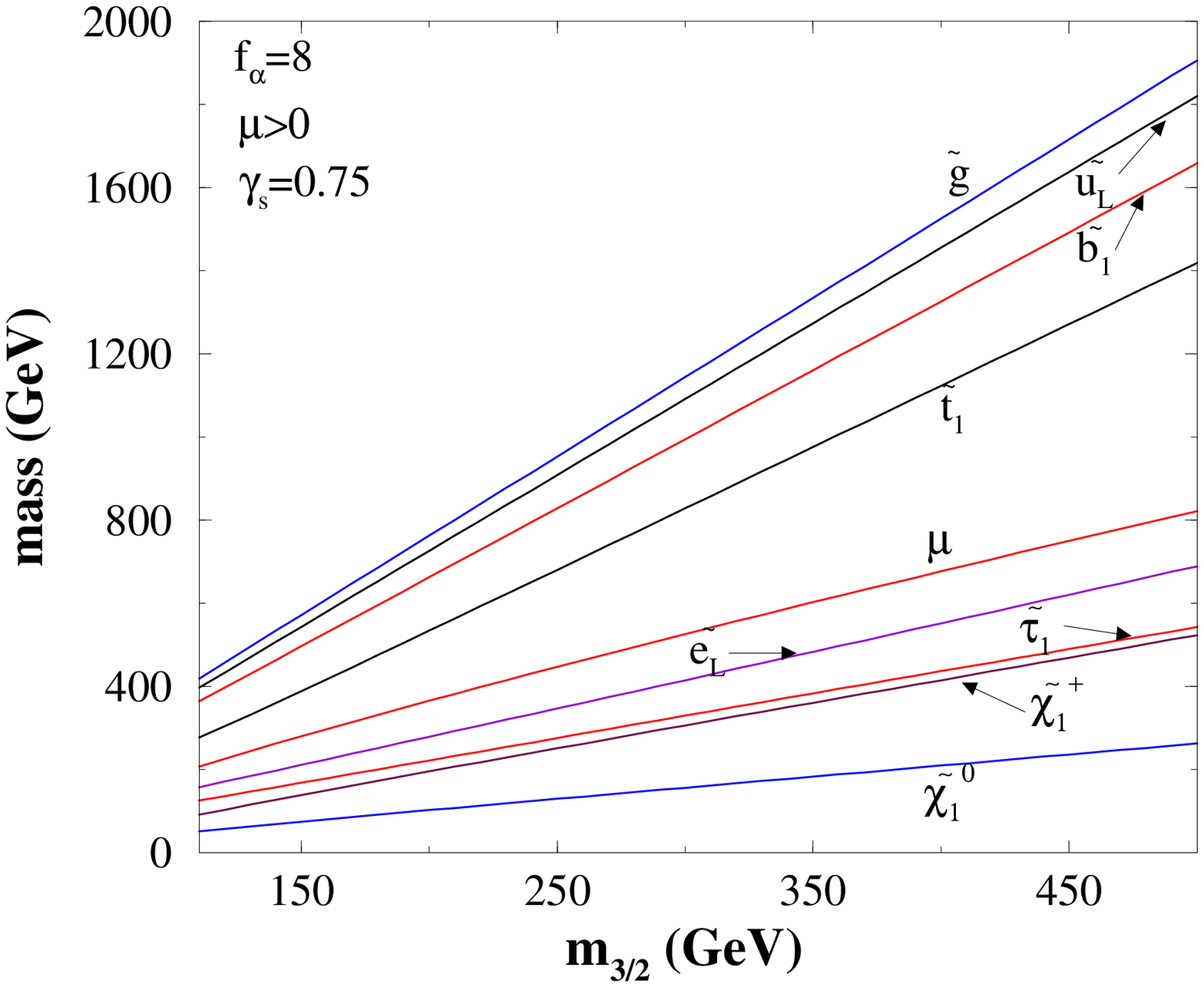} 
\end{minipage}}
\caption{}                       
\label{gamma_mass_gravi_mass}             
\end{figure}

A plot of the mass spectrum as a function of $\gamma_s$ 
 with determined $\tan\beta$ but without the
imposition of the relic density constraint is given in  
Fig.~\ref{gamma_vs_mass}.
A similar analysis as a function of $\mgravi$ is given in 
Fig.~\ref{gravi_vs_mass}
but again without the imposition of the relic density constraints.
Inclusion of the relic density constraint limits $\mgravi$ to lie 
lower than about 500 GeV and the corresponding upper limits
on sparticle masses can be read off from  Fig.~\ref{gravi_vs_mass}. 
In this case
the upper limits on sparticle masses all lie roughly below 2 TeV and
all are essentially accessible at the LHC. 
We note that  the gluino is 
the highest mass particle over most of the allowed parameter space
of the model. Inclusion of the FCNC constraint 
drastically lowers the upper limits so that the gluino mass lies below 
$500$ GeV and masses of the other particles are even 
lower.  Thus some of the particles may also be accessible at the Tevatron.
Another interesting phenomenon is that over  most of the allowed parameter
space the neutralino is the lowest mass supersymmetric particle (LSP)
and thus with R parity a candidate for cold dark matter.

\begin{figure}                       
\vspace*{-0.6in}                                 
\subfigure[ Scatter plot for spin 
independent LSP-proton cross section vs LSP mass for 
$\mu>0$ when $\gamma_s$ and $\mgravi$ are scanned. 
The region with black circles satisfies the WMAP constraint. Present 
limits (top three contours) and 
future accessibility regions are shown.]{
\label{sigma_si}             
\hspace*{-0.4in}                               
\begin{minipage}[b]{\textwidth}                       
\centering                      
\includegraphics[width=0.9\textwidth, height=0.45\textwidth]{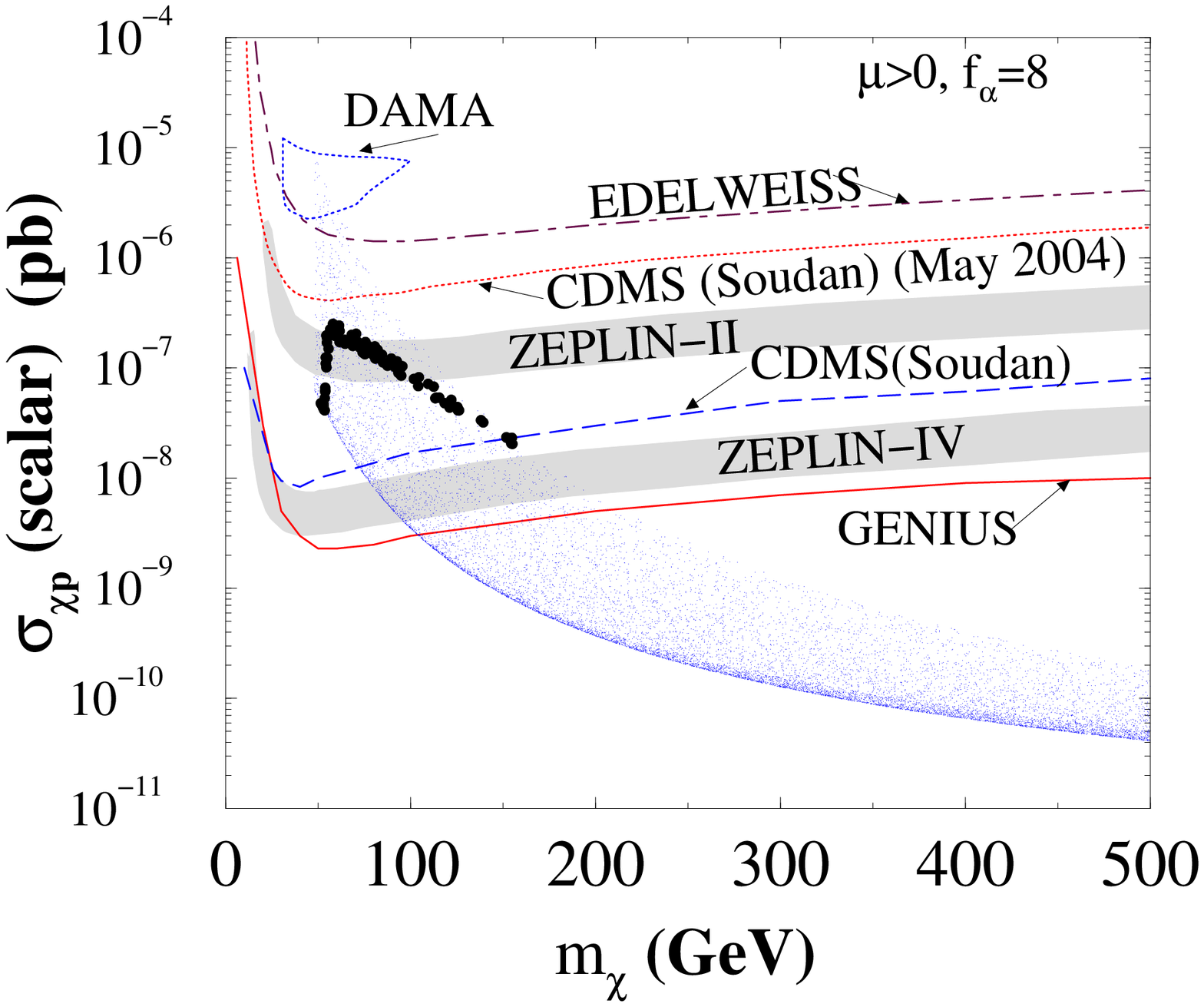} 
\vspace{2mm}                                 
\end{minipage}}                       

\subfigure[Spin-dependent LSP-proton cross section vs LSP mass for $\mu>0$. 
The region with black circles satisfies the WMAP constraint.]{
\label{sigma_sd}             
\hspace*{-0.4in}                               
\begin{minipage}[b]{\textwidth}                       
\centering                      
\includegraphics[width=0.9\textwidth, height=0.45\textwidth]{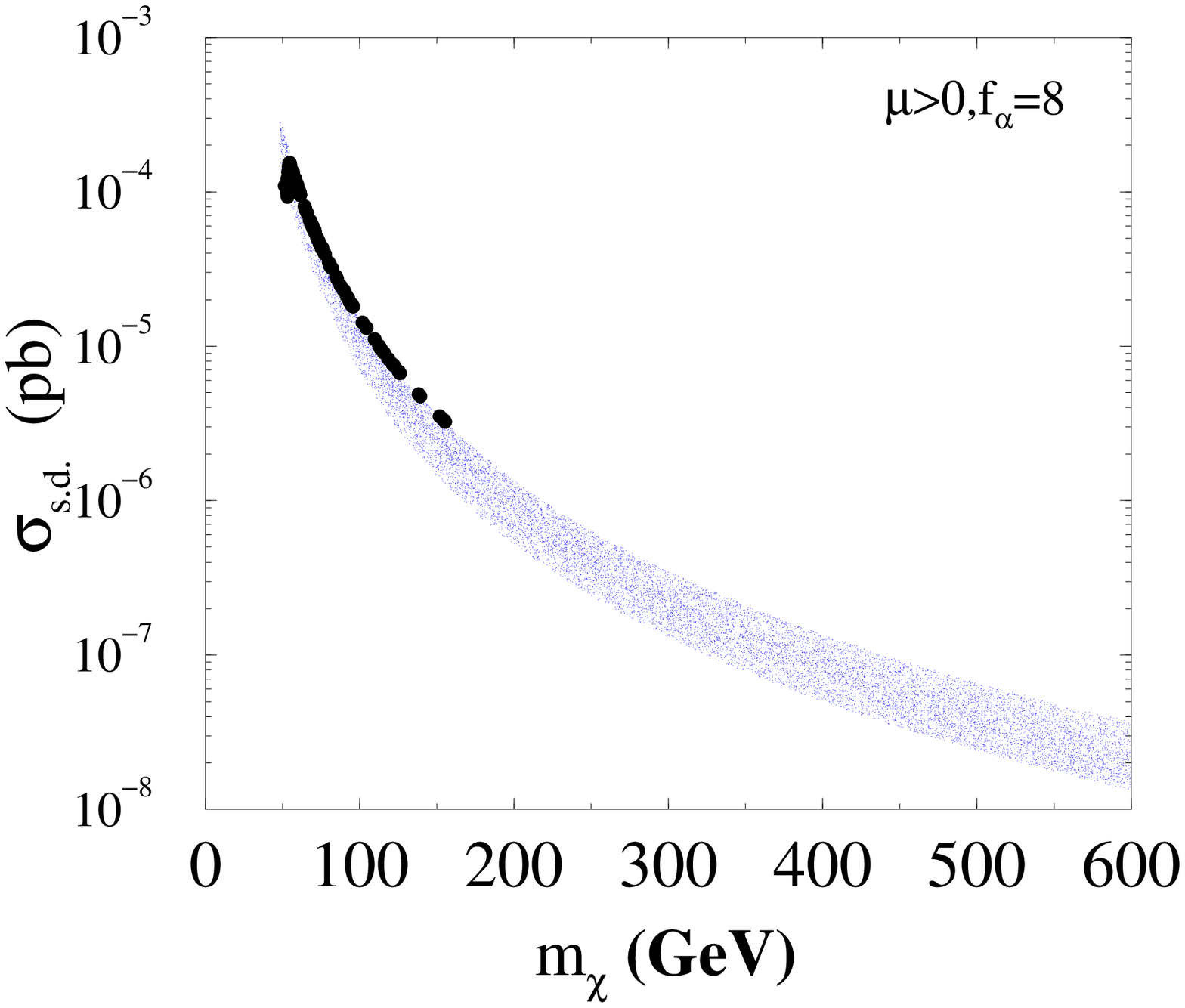} 
\vspace{2mm}                                 
\end{minipage}}
\caption{}
\label{plus_sigma}
\end{figure}

In Fig.~\ref{sigma_si} 
we give a plot of the neutralino-proton scalar 
cross section  $\sigma_{\chi p}$  as a function of the neutralino mass for 
$\mu>0$.
 The region with black circles satisfies the WMAP constraint.
Also exhibited are the sensitivities that will be
reached by experiment, i.e.,  CDMS (Soudan)\cite{cdms},  
GENIUS\cite{genius} and ZEPLIN\cite{cline,smith} 
(shown as two broad bands). In Fig.~\ref{sigma_si} 
we have included the limit from the first set of data from   
CDMS(Soudan) as announced recently\cite{cdmsmay2004} as well as 
 the latest contour from EDELWEISS\cite{edelweiss}. 
 The dotted enclosed region is the contour from DAMA\cite{damaresult}.
The analysis shows that in this class of string models dark matter  
falls within the
sensitivity that will be achievable at the CDMS (Soudan), GENIUS and ZEPLIN.
A similar analysis for the spin dependent neutralino-proton 
cross section  $\sigma_{\chi p}$ as a function of the neutralino mass
is given in  Fig.~\ref{sigma_sd} for $\mu>0$.  The region with black circles 
satisfies the WMAP constraint.

\begin{figure}                       
\vspace*{-0.6in}                                 
\subfigure[
Contours of constant $\tan\beta$, and 
$\mu$ are shown on the available parameter space 
for $\mu<0$ in the $(\gamma_s-m_{3/2})$ plane. 
The $b \rightarrow s + \gamma$ contour 
is shown as a dot-dashed line  below  which the region is 
disallowed. WMAP satisfied relic density region is shown as small 
shaded area in black. The disallowed gray region I and III are 
typically discarded because of the absence of consistent GUT scale 
inputs. The region II typically refers to absence of REWSB or smaller 
than experimental lower limits of $m_{\tilde \chi_1^\pm}$. 
The region IV is a no solution zone like I and 
III, but its location and extent depends on the sensitivity of 
the minimization scale for REWSB. Region V is the tachyonic 
$\tilde \tau_1$ zone. 
]{
\label{min_gamma_gravi}             
\hspace*{-0.4in}                               
\begin{minipage}[b]{\textwidth}                       
\centering                      
\includegraphics[width=0.8\textwidth, height=0.4\textwidth]{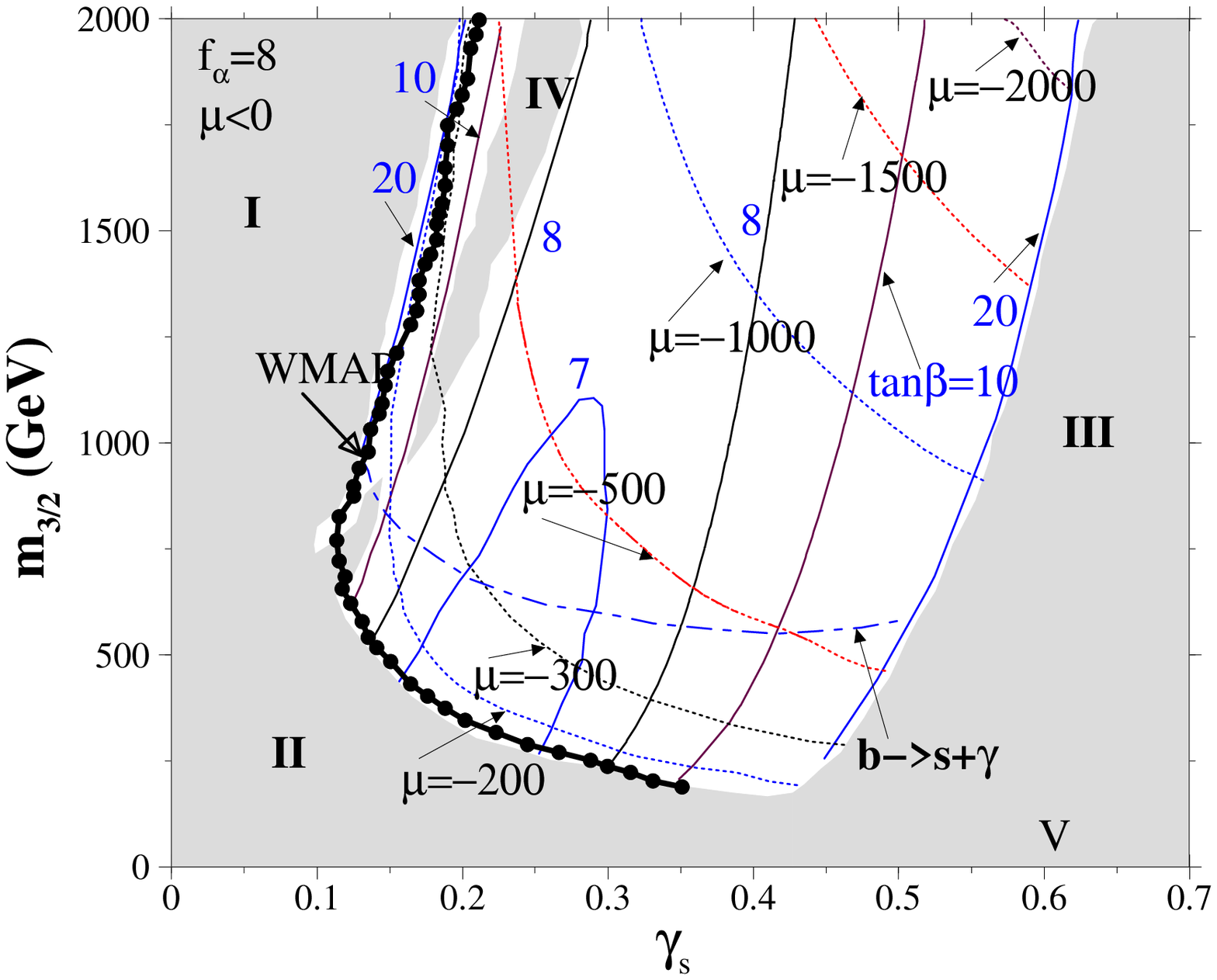} 
\end{minipage}}                       

\subfigure[
The contours of constant $m_{3/2}$ in the 
$\gamma_s-\tan\beta$ plane for $\mu<0$ are as labelled. The shaded (cyan) 
regions satisfy the $b \rightarrow s + \gamma$ limits. The black dotted  
area does not satisfy $b \rightarrow s + \gamma$ limits.
The WMAP allowed region is shown in filled circles in 
black.]{
\label{min_gamma_tan}             
\hspace*{-0.4in}                               
\begin{minipage}[b]{\textwidth}                       
\centering                      
\includegraphics[width=0.8\textwidth, height=0.4\textwidth]{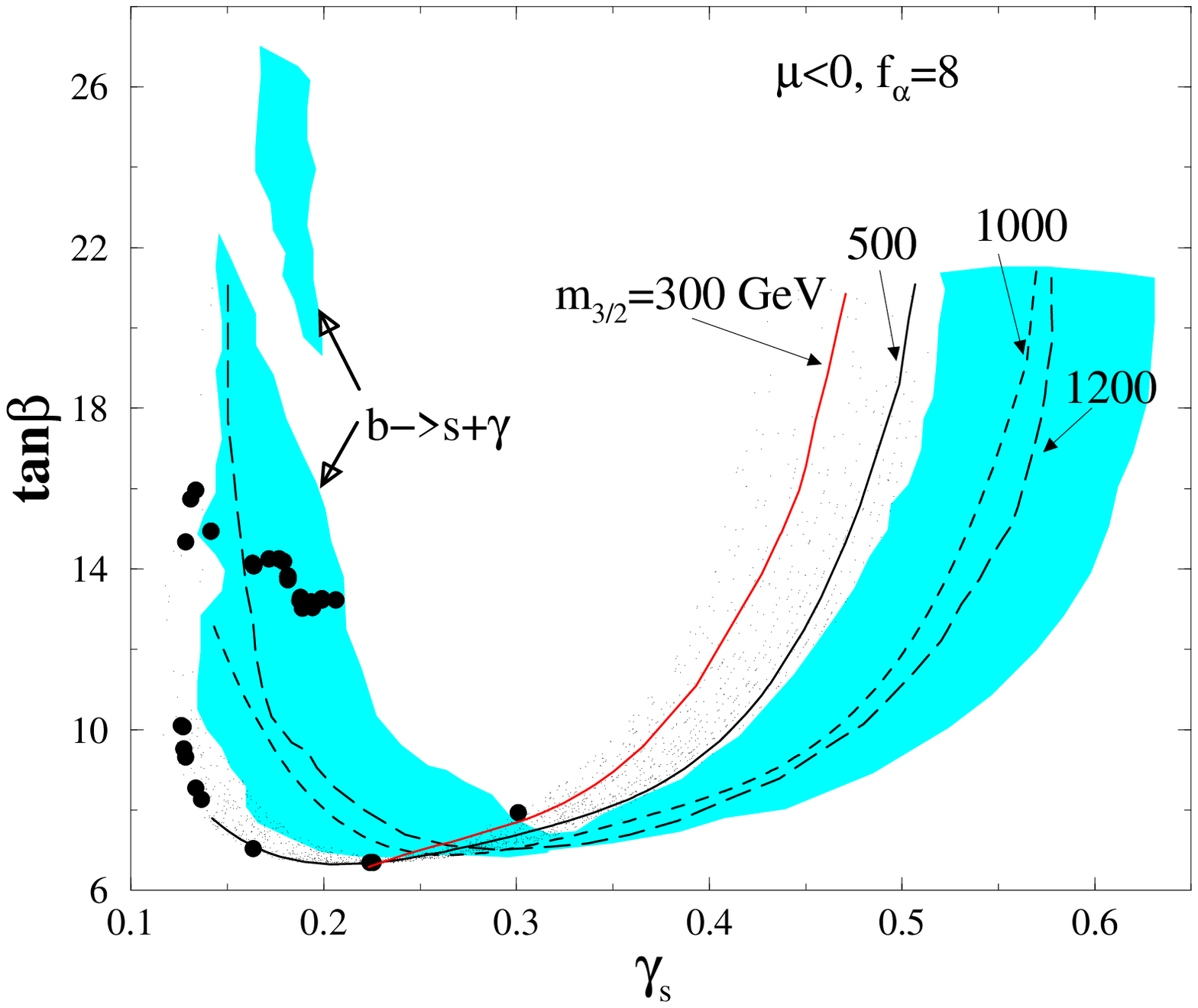} 
\end{minipage}}
\caption{}        
\label{min_gamma_vs_gravi_tan}                                
\end{figure}

\begin{figure}                       
\vspace*{-0.6in}                                 
\subfigure[Variation of sparticle masses with respect to 
$\gamma_s$ when $m_{3/2}$ is fixed at 1000 GeV for $\mu<0$.
The WMAP constraint is not exhibited.]{
\label{min_gamma_vs_mass}             
\hspace*{-0.4in}                               
\begin{minipage}[b]{\textwidth}                       
\centering                      
\includegraphics[width=0.9\textwidth, height=0.5\textwidth]{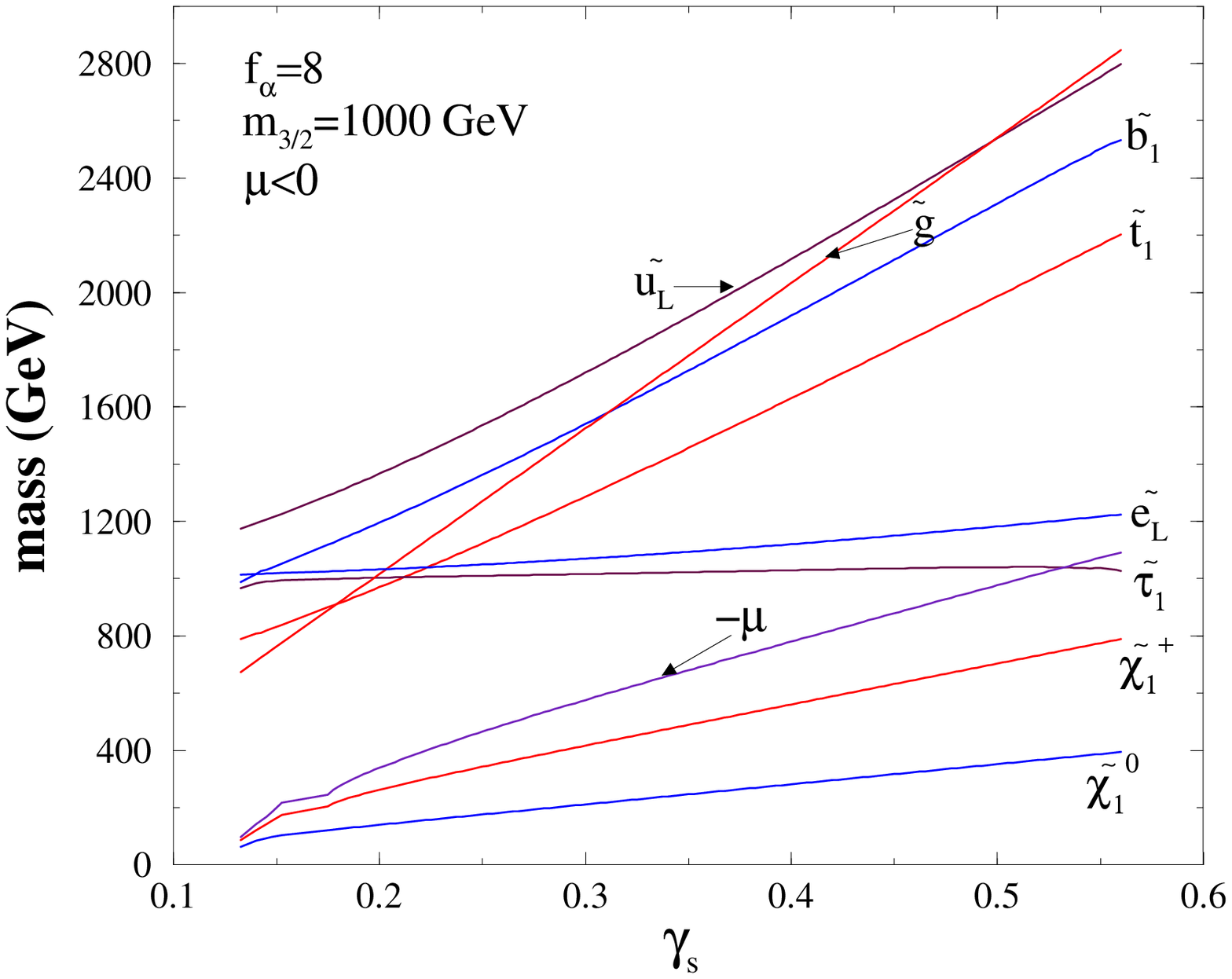} 
\end{minipage}}                       

\subfigure[
Variation of sparticle masses with respect to 
$m_{3/2}$ when $\gamma_s$ is fixed at 0.35 for $\mu<0$.
The WMAP constraint is not exhibited.]{
\label{min_gravi_vs_mass}             
\hspace*{-0.4in}                               
\begin{minipage}[b]{\textwidth}                       
\centering                      
\includegraphics[width=0.9\textwidth, height=0.5\textwidth]{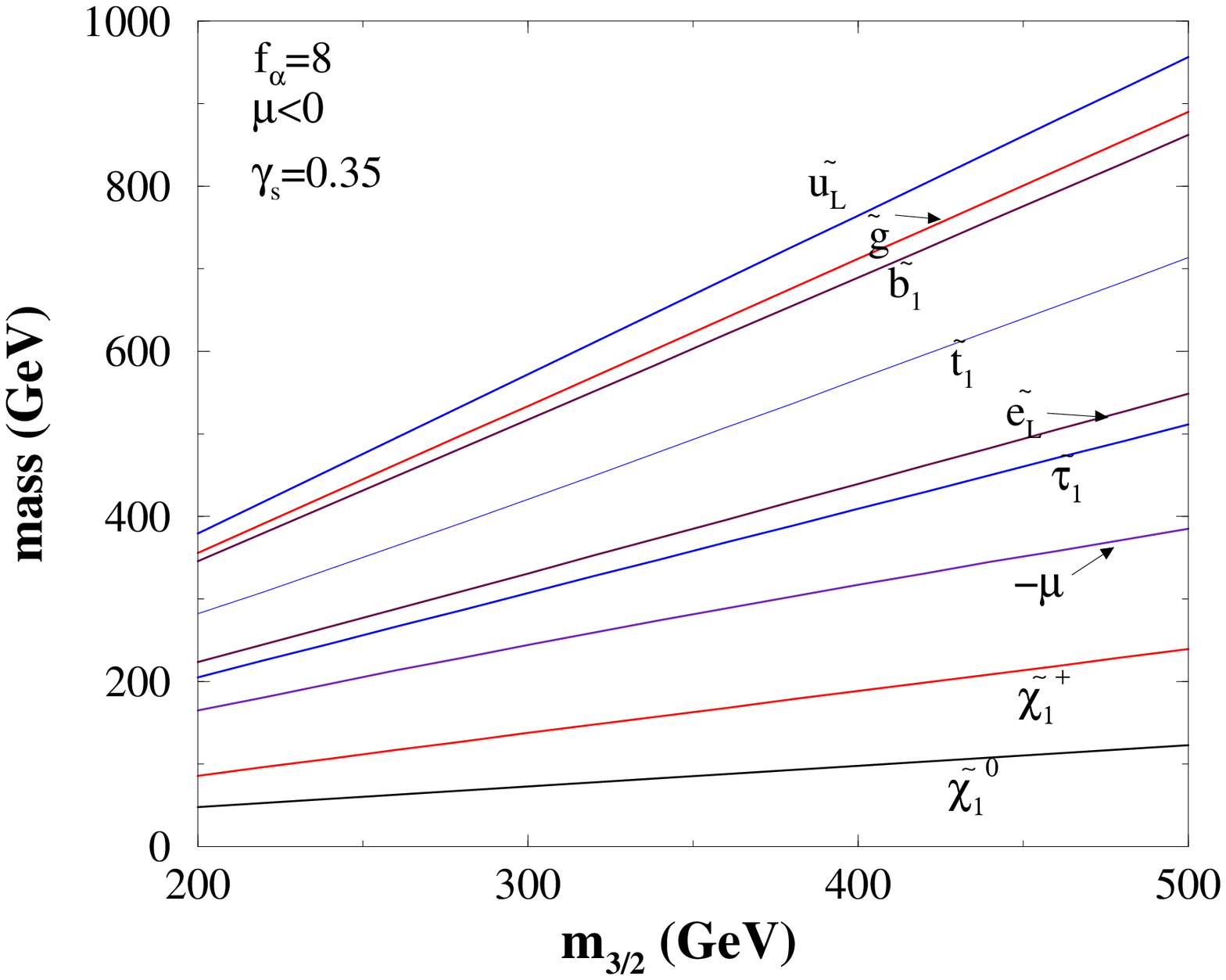}  
\end{minipage}}
\caption{}                       
\label{min_gamma_mass_gravi_mass}             
\end{figure}

\begin{figure}                       
\vspace*{-0.6in}                                 
\subfigure[
Scatter plot for spin 
independent LSP-proton cross section vs LSP mass for 
$\mu<0$ when $\gamma_s$ and $\mgravi$ are scanned. 
The region with black circles satisfies the WMAP constraint. Present 
limits (top three contours) and future accessibility regions are shown.
]{
\label{min_sigma_si}             
\hspace*{-0.4in}                               
\begin{minipage}[b]{\textwidth}                       
\centering                      
\includegraphics[width=0.9\textwidth, height=0.45\textwidth]{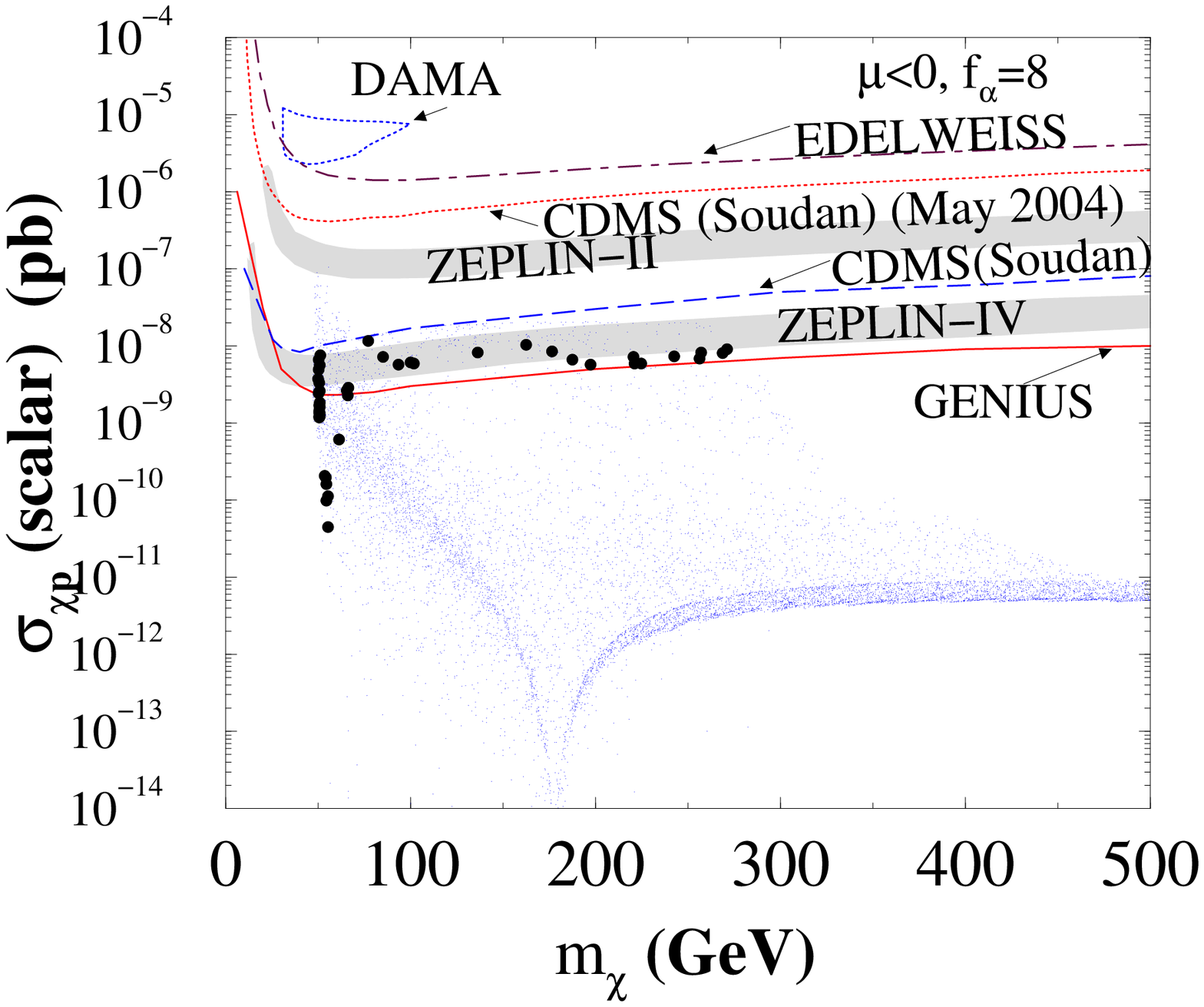} 
\vspace{2mm}
\end{minipage}}                       

\subfigure[Spin-dependent LSP-proton cross section vs LSP mass for $\mu<0$. 
The region with black circles satisfies the WMAP constraint.]{
\label{min_sigma_sd}             
\hspace*{-0.4in}                               
\begin{minipage}[b]{\textwidth}                       
\centering                      
\includegraphics[width=0.9\textwidth, height=0.45\textwidth]{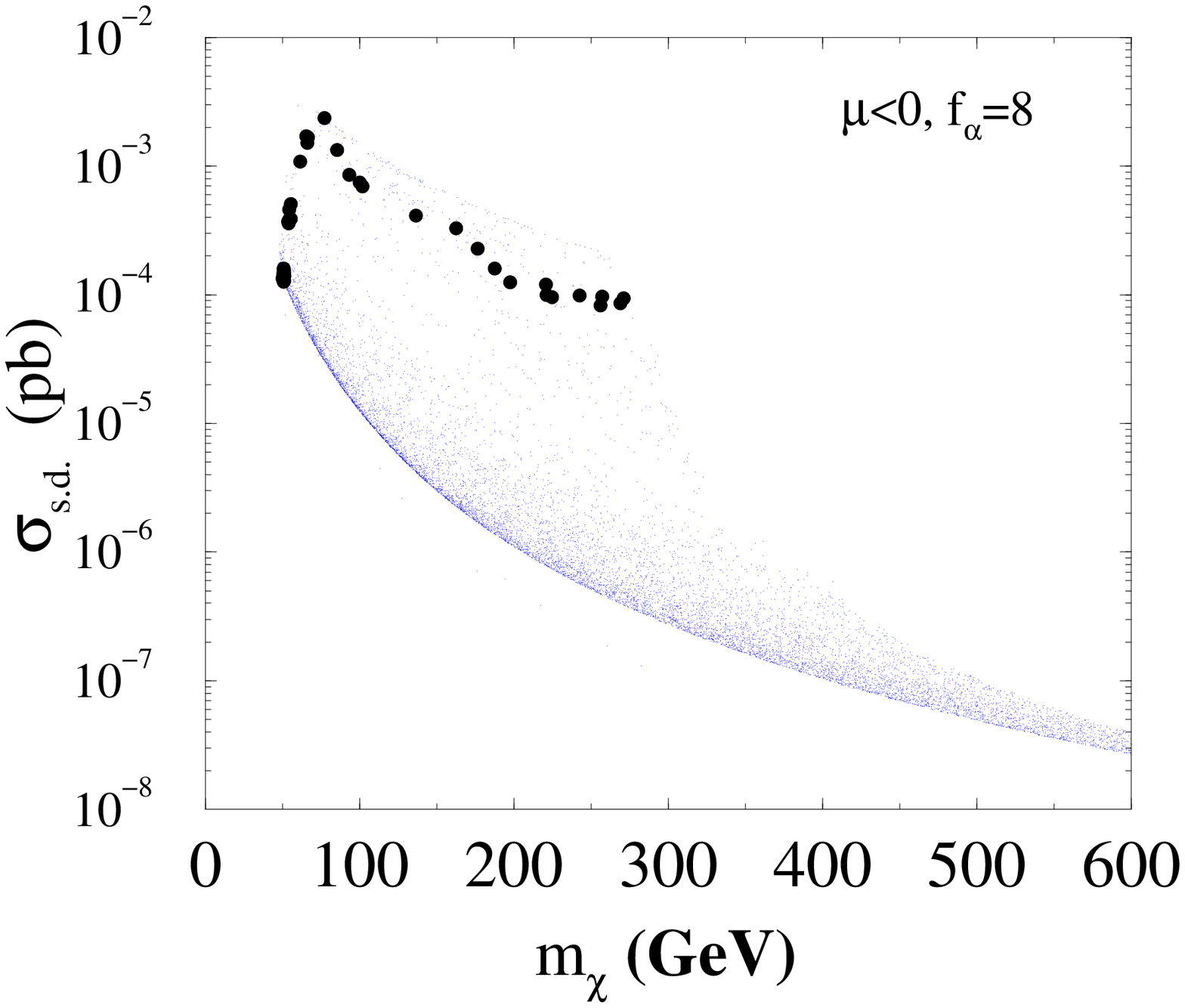} 
\vspace{2mm}
\end{minipage}}
\caption{}
\label{min_sigma}
\end{figure}

  The $\mu<0$ case is analyzed in 
Figs.~(\ref{min_gamma_vs_gravi_tan})-(\ref{min_sigma}).
In Fig.~(\ref{min_gamma_vs_gravi_tan})
contours of $\tan\beta$, and 
$\mu$ are shown in the $\gamma_s-m_{3/2}$ plane. 
The constraint from $b \rightarrow s + \gamma$  
is shown as a dot-dashed line  below  which the region is 
disallowed.  The WMAP satisfied relic density region is shown as a small 
shaded area in black. The disallowed gray region I and III is 
discarded  typically due to the absence of 
a $\tan\beta$ solution consistent with GUT 
scale inputs. The region  II  typically refers to the absence of REWSB, or 
having smaller than experimental lower limits of $m_{\tilde \chi_1^\pm}$.
The region IV is a no solution zone like I and 
III, but its location and extent depends on the sensitivity of 
the minimization scale for REWSB.  The region V is excluded because of  
tachyonic $\tilde \tau_1$. 
An important point to note for $\mu<0$ 
case is that a given $\tan\beta$ contour has two separate regions, 
corresponding to smaller and larger $\gamma_s$ values for 
a given $m_{3/2}$.  Furthermore, $\tan\beta$ is severely 
constrained to lie below 20 for $\mu<0$. 
Unlike the $\mu>0$ case, the $\mu<0$ branch  
has a larger region where $|\mu|$ is very small, close to its value 
constrained by the lighter chargino mass lower limit. 
Further, $m_{3/2}$ here can acquire values much larger than
in the  $\mu>0$  case consistent with the WMAP constraint
 and the REWSB  appears to be occurring on the hyperbolic branch.
Fig.~\ref{min_gamma_tan} shows contours of $m_{3/2}$ in  
$\gamma_s-\tan\beta$ plane for $\mu<0$.   The lightly 
shaded areas (in cyan) satisfy the $b \rightarrow s + \gamma$ limits.  
In contrast, areas marked 
with black dots do not satisfy the $b \rightarrow s + \gamma$ bounds.  
The WMAP allowed region is shown in small filled circles in 
black.  Clearly, the WMAP constraint along with $b \rightarrow s + \gamma$ 
restricts $\tan\beta$ to be within $12$ to $15$ .
A plot of the mass spectrum as a function of $\gamma_s$ 
is given in Fig.~\ref{min_gamma_vs_mass}.
A similar analysis as a function of $\mgravi$ is given in 
Fig.~\ref{min_gravi_vs_mass}.
Fig.~\ref{min_sigma_si} shows 
a plot of the neutralino-proton scalar 
cross section  $\sigma_{\chi p}$ vs neutralino mass for 
$\mu<0$.  Here the detection cross sections are smaller than the 
case for $\mu>0$. However, the values of spin-dependent cross section 
of Fig.~\ref{min_sigma_sd} for $\mu<0$ are much higher than 
the case when $\mu>0$.  

\begin{figure}                       
\vspace*{-0.6in}                                 
\subfigure[
Analysis at the self dual point $f_{\alpha}=3\sqrt 3$ where
contours of constant $\tan\beta$ are shown on the available 
parameter space for $\mu>0$ in the $(\gamma_s-m_{3/2})$ plane. 
The $b \rightarrow s + \gamma$ contour 
is shown as a dot-dashed line below which the region is 
disallowed. WMAP satisfied relic density region is shown as small 
shaded area in black. The gray region-I refers to the  discarded 
zone of very large $\tan\beta$ where 
Yukawa couplings are beyond the perturbative domain.
The gray region II results from 
the absence of REWSB or smaller $m_{\tilde \chi_1^\pm}$ below 
the experimental limit.  
]{
\label{gamma_gravi_fnew}             
\hspace*{-0.4in}                               
\begin{minipage}[b]{\textwidth}                       
\centering                      
\includegraphics[width=0.8\textwidth, height=0.4\textwidth]{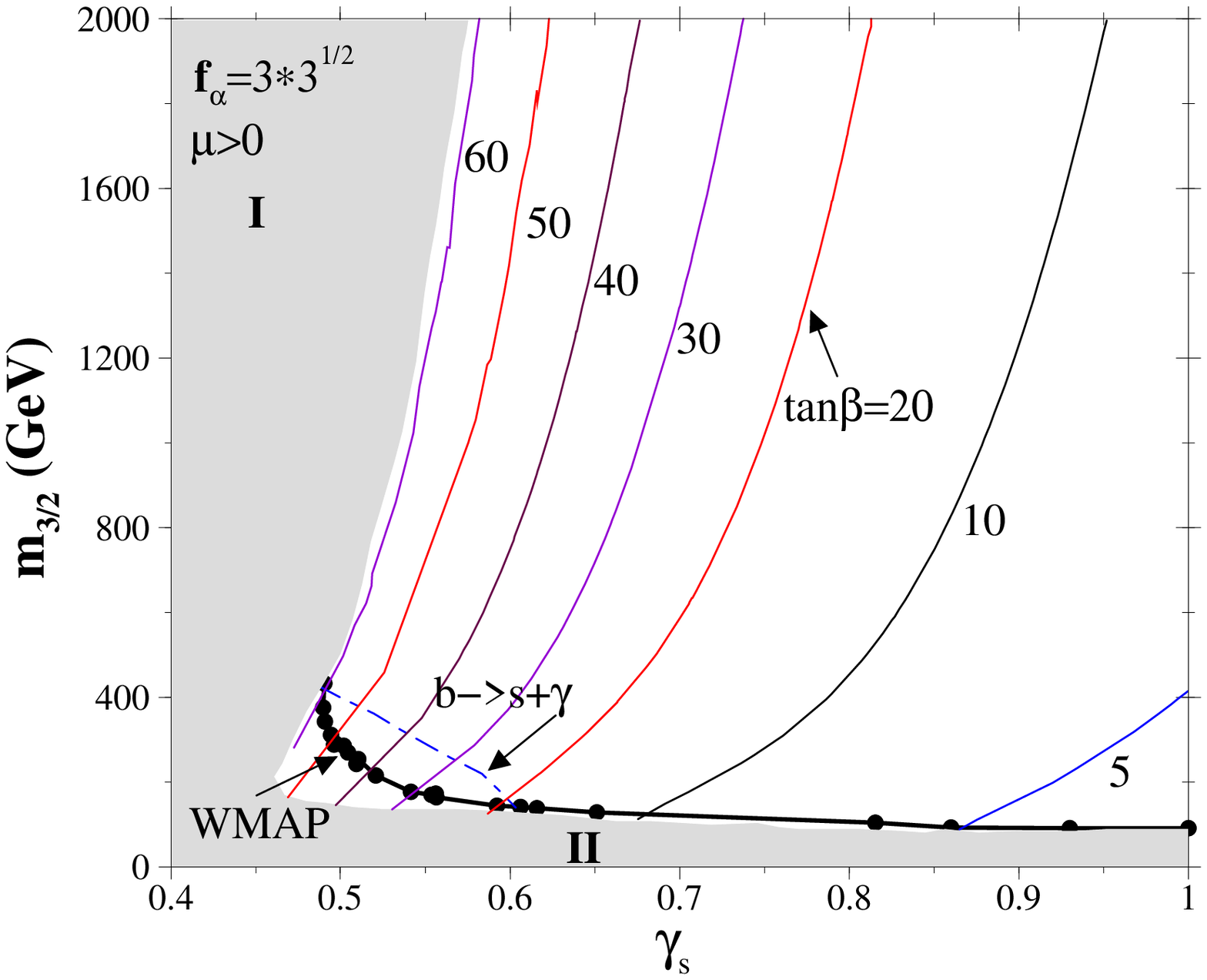} 
\end{minipage}}                       

\subfigure[
Analysis at the self dual point $f_{\alpha}=3\sqrt 3$ where
a scatter plot for spin 
independent LSP-proton cross section vs LSP mass for 
$\mu>0$ is shown when $\gamma_s$ and $\mgravi$ are scanned. 
The region with black circles satisfies the WMAP constraint. Present 
limits (top three contours) and 
future accessibility regions are shown.]{
\label{sigma_si_fnew}             
\hspace*{-0.4in}                               
\begin{minipage}[b]{\textwidth}                       
\centering                      
\includegraphics[width=0.8\textwidth, height=0.4\textwidth]{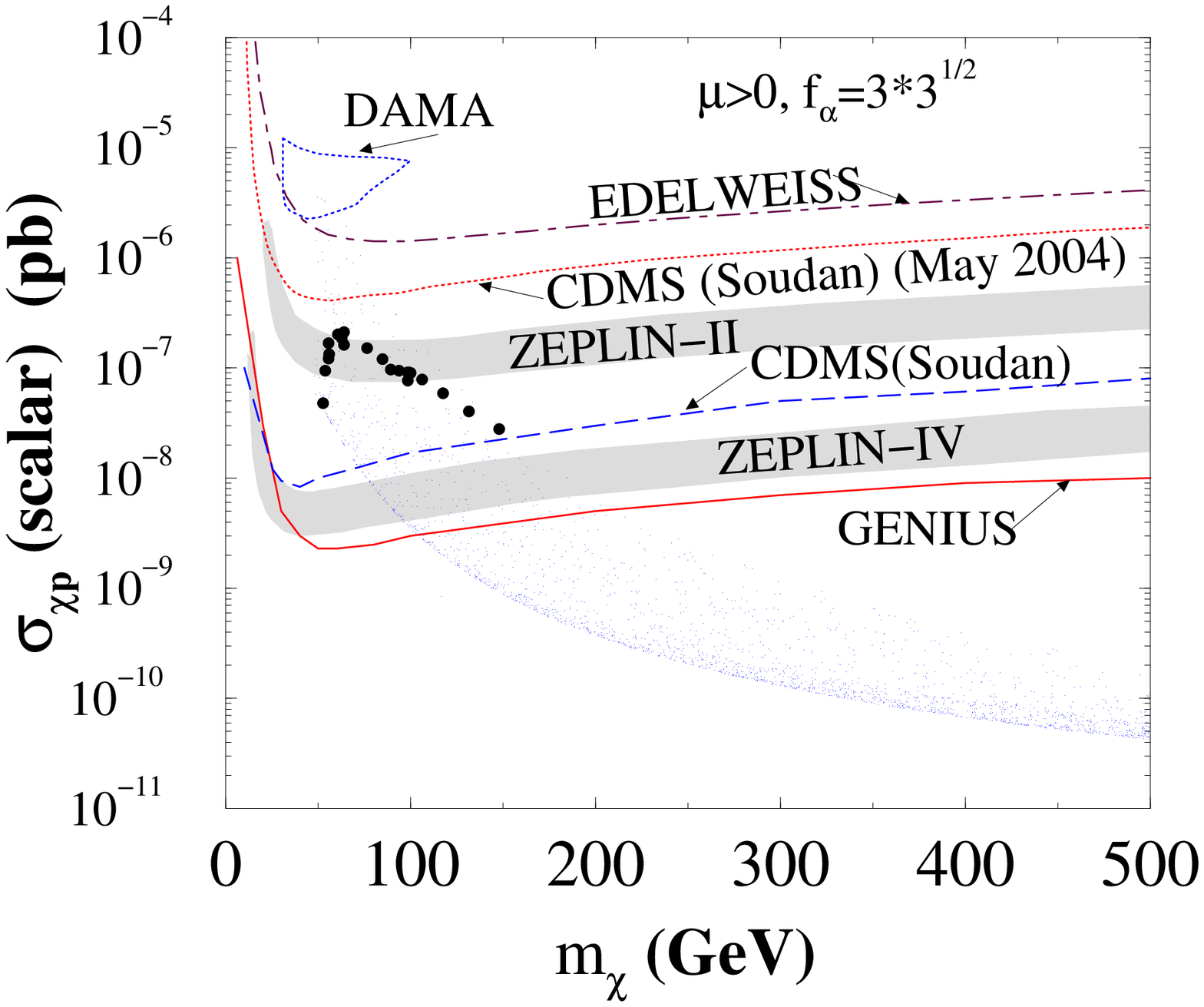} 
\vspace{2mm}
\end{minipage}}
\caption{}    
\label{selfdual_3sq3}                   
\end{figure}

\begin{figure}                       
\vspace*{-0.6in}                                 
\subfigure[
Analysis at the self dual point $f_{\alpha}=3\sqrt 3$ where
contours of constant $\tan\beta$ are shown on the available 
parameter space for $\mu<0$ in the $(\gamma_s-m_{3/2})$ plane. 
The $b \rightarrow s + \gamma$ contour 
is shown as a dot-dashed line  below  which the region is 
disallowed. WMAP satisfied relic density region is shown as small 
shaded area in black. 
The disallowed regions I to V are same as those in 
Fig.(\ref{min_gamma_gravi}). 
]{
\label{min_gamma_gravi_fnew}             
\hspace*{-0.4in}                               
\begin{minipage}[b]{\textwidth}                       
\centering                      
\includegraphics[width=0.8\textwidth, height=0.4\textwidth]{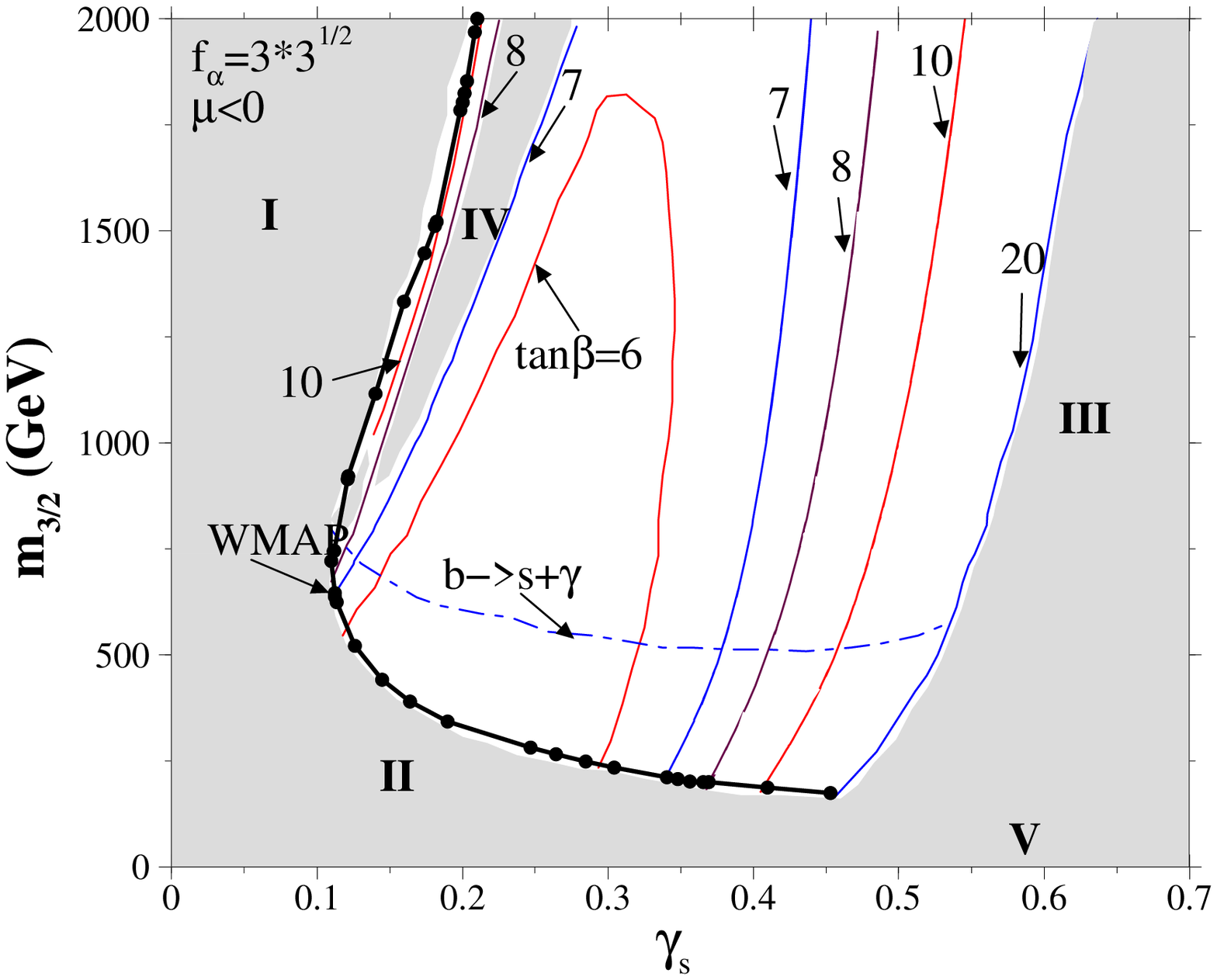} 
\end{minipage}}                       

\subfigure[
Analysis at the self dual point $f_{\alpha}=3\sqrt 3$ where
a scatter plot for spin 
independent LSP-proton cross section vs LSP mass for 
$\mu<0$ is shown when $\gamma_s$ and $\mgravi$ are scanned. 
The region with black circles satisfies the WMAP constraint. Present 
limits (top three contours) and 
future accessibility regions are shown.]{
\label{min_sigma_si_fnew}             
\hspace*{-0.4in}                               
\begin{minipage}[b]{\textwidth}                       
\centering                      
\includegraphics[width=0.8\textwidth, height=0.4\textwidth]{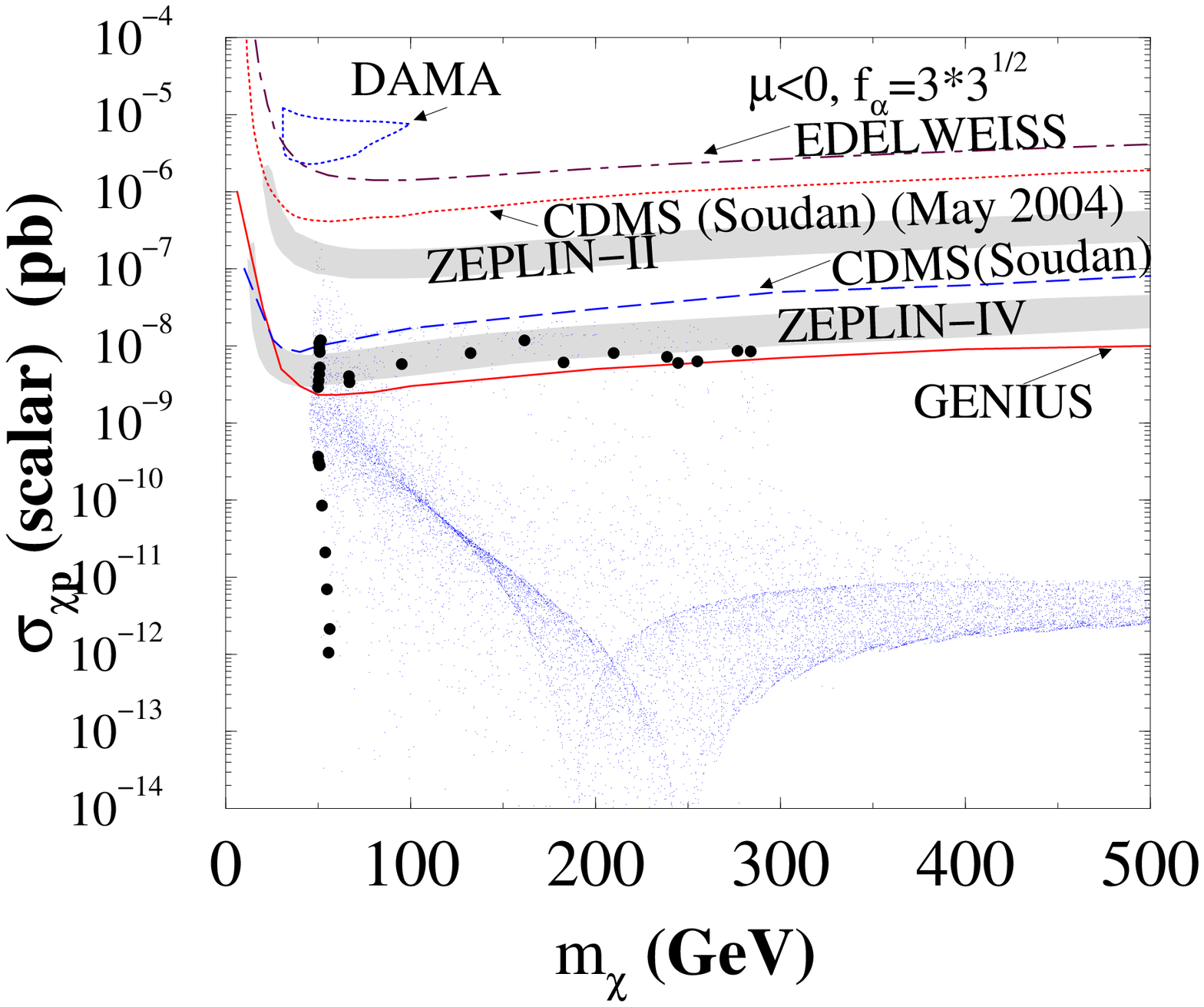} 
\vspace{2mm}
\end{minipage}}
\caption{}    
\label{min_selfdual_3sq3}                   
\end{figure}

We discuss now the dependence of the analysis on the choice of the self dual
point. In Figs.~(\ref{selfdual_3sq3}) and (\ref{min_selfdual_3sq3}) 
we give the analysis for the case when $f_{\alpha}= 3\sqrt 3$.
A comparison of Fig.~(\ref{selfdual_3sq3}) with
Fig.~(\ref{gamma_gravi}) and with Fig.~(\ref{sigma_si}) for the 
$\mu>0$ case shows that the analysis at the self dual point 
$f_{\alpha}= 3\sqrt 3$ follows a similar pattern as the  analysis at
the self dual point $f_{\alpha}= 8$. The main difference is the lowering
of the maximum in $m_{\frac{3}{2}}$ by about a 100 GeV without imposition
of the $b\rightarrow s+\gamma$ constraint.  However, when   
$b\rightarrow s+\gamma$ constraint is imposed there is no significant 
change with respect to the result corresponding to $f_\alpha=8$. 
For the $\mu<0$ case a comparison of Figs.~(\ref{min_selfdual_3sq3})
 with Fig.~(\ref{min_gamma_gravi}) and with Fig.~(\ref{min_sigma_si})
shows that the  spin independent $\sigma_{\chi p}$ cross section can
dip lower for the case   $f_{\alpha}= 3\sqrt 3$ compared to the
case when $f_{\alpha}= 8$. Aside from that the features of
 Figs.~(\ref{min_selfdual_3sq3}) are very similar to those of 
Fig.~(\ref{min_gamma_gravi}) and with Fig.~(\ref{min_sigma_si}).

\begin{figure}                       
\vspace*{-0.6in}                                 
\subfigure[
Composite analysis for relic density and $Br(b\rightarrow s+\gamma)$ 
in $(\gamma_s - \mgravi)$ plane for $\mu>0$ with the inclusion of $U_i$ moduli 
at the self dual points corresponding to 
all possible values of $f_\alpha: 2^n3^{3-n/2},n=0,...6$. 
The $b \rightarrow s + \gamma$ contour 
is shown as a dot-dashed line below which the region is maximally  
disallowed.  WMAP satisfied relic density region is maximally shown as small 
shaded area in black. The gray areas are similar to the 
discarded regions of Fig.(\ref{gamma_gravi}).
]{
\label{pos_relic_umod}             
\hspace*{-0.4in}                               
\begin{minipage}[b]{\textwidth}                       
\centering                      
\includegraphics[width=0.8\textwidth, height=0.4\textwidth]{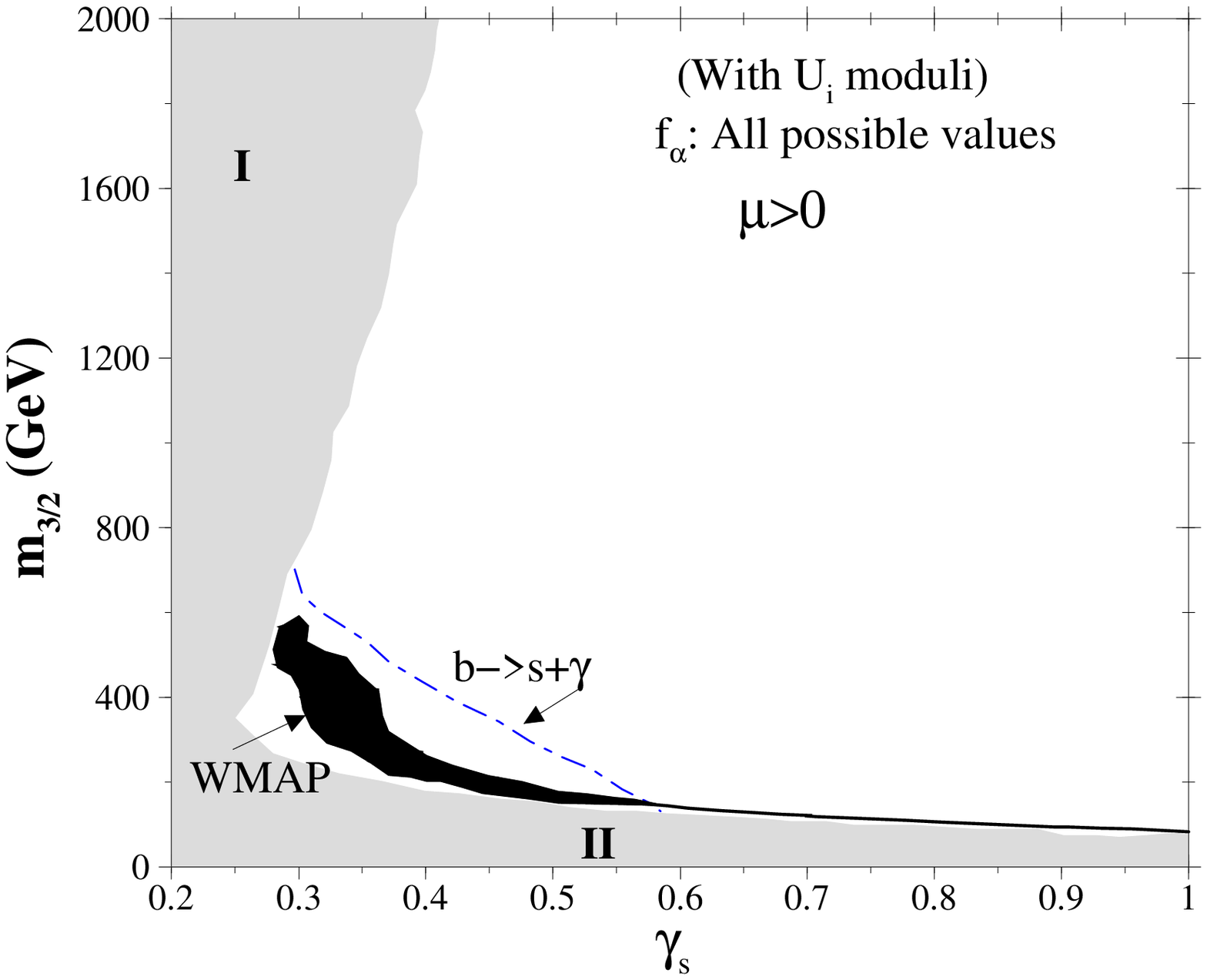}
\end{minipage}}                       

\subfigure[
Scatter plot for spin 
independent LSP-proton cross section vs LSP mass for 
$\mu>0$ when $\gamma_s$ and $\mgravi$ are scanned, for the composite analysis 
with $U_i$ moduli corresponding to Fig.(\ref{pos_relic_umod}). 
The region with black circles satisfies the WMAP constraint. 
]{
\label{pos_sigmasi_umod}             
\hspace*{-0.4in}                               
\begin{minipage}[b]{\textwidth}                       
\centering                      
\includegraphics[width=0.8\textwidth, height=0.4\textwidth]{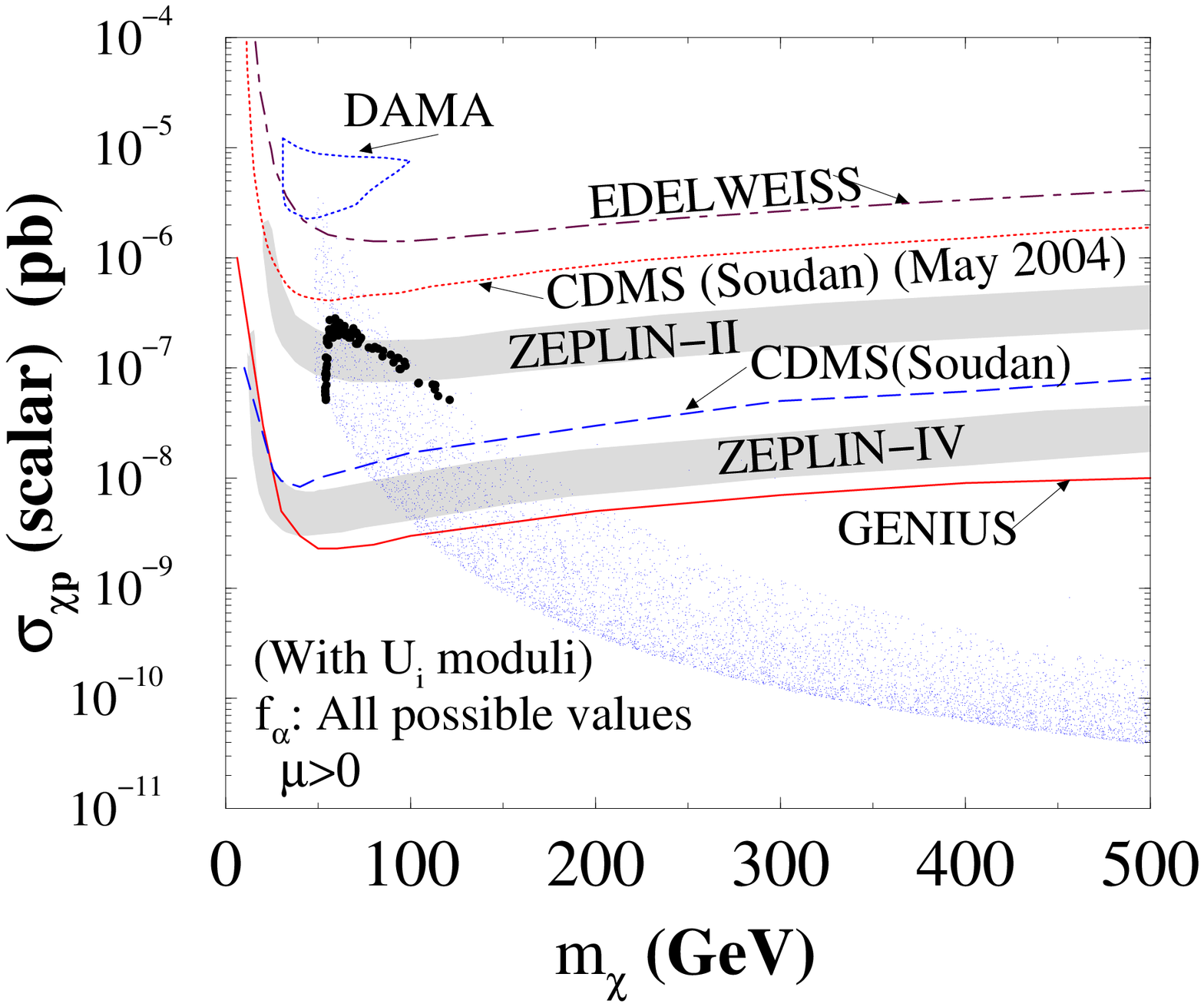} 
\vspace{2mm}
\end{minipage}}
\caption{}    
\label{pos_umod}                   
\end{figure}

\begin{figure}                       
\vspace*{-0.6in}                                 
\subfigure[
Composite analysis for relic density and $Br(b\rightarrow s+\gamma)$ 
in $(\gamma_s - \mgravi)$ plane for $\mu<0$ with the inclusion of $U_i$ moduli 
at the self dual points corresponding to 
all possible values of $f_\alpha: 2^n3^{3-n/2},n=0,...6$. 
The $b \rightarrow s + \gamma$ contour 
is shown as a dot-dashed line below which the region is maximally  
disallowed.  WMAP satisfied relic density region is maximally shown as small 
shaded area in black. The gray areas are similar to the 
discarded regions of Fig.(\ref{min_gamma_gravi}).
]{
\label{neg_relic_umod}             
\hspace*{-0.4in}                               
\begin{minipage}[b]{\textwidth}                       
\centering                      
\includegraphics[width=0.8\textwidth, height=0.4\textwidth]{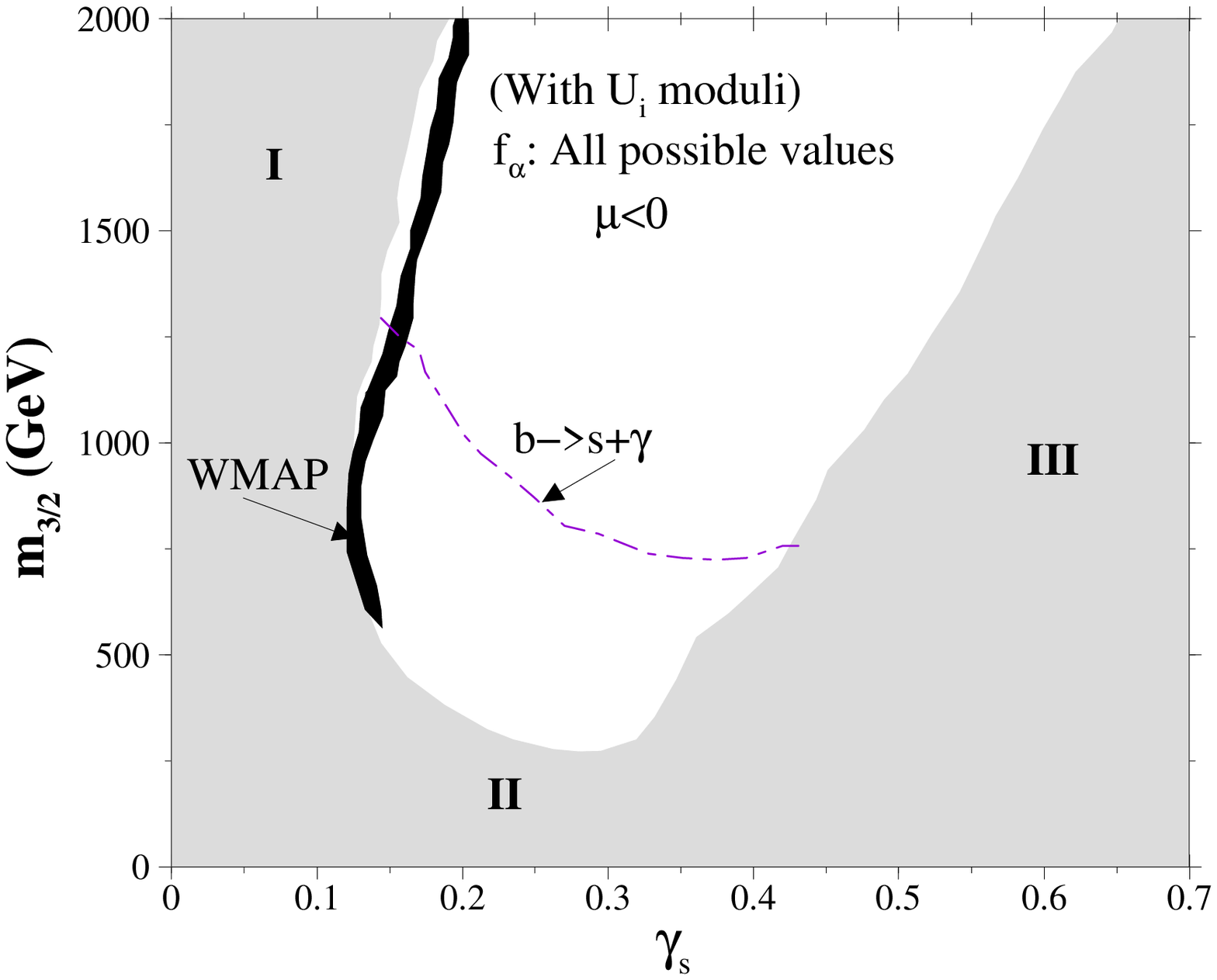}
\end{minipage}}                       

\subfigure[
Scatter plot for spin 
independent LSP-proton cross section vs LSP mass for 
$\mu<0$ when $\gamma_s$ and $\mgravi$ are scanned, for the composite analysis 
with $U_i$ moduli corresponding to Fig.(\ref{neg_relic_umod}). 
A few scattered points shown in black circles satisfies the WMAP 
constraint. Most of such points corresponds to 
$\mx1>>500$~GeV, thus falling outside the displayed 
limit.  
]{
\label{neg_sigmasi_umod}             
\hspace*{-0.4in}                               
\begin{minipage}[b]{\textwidth}                       
\centering                      
\includegraphics[width=0.8\textwidth, height=0.4\textwidth]{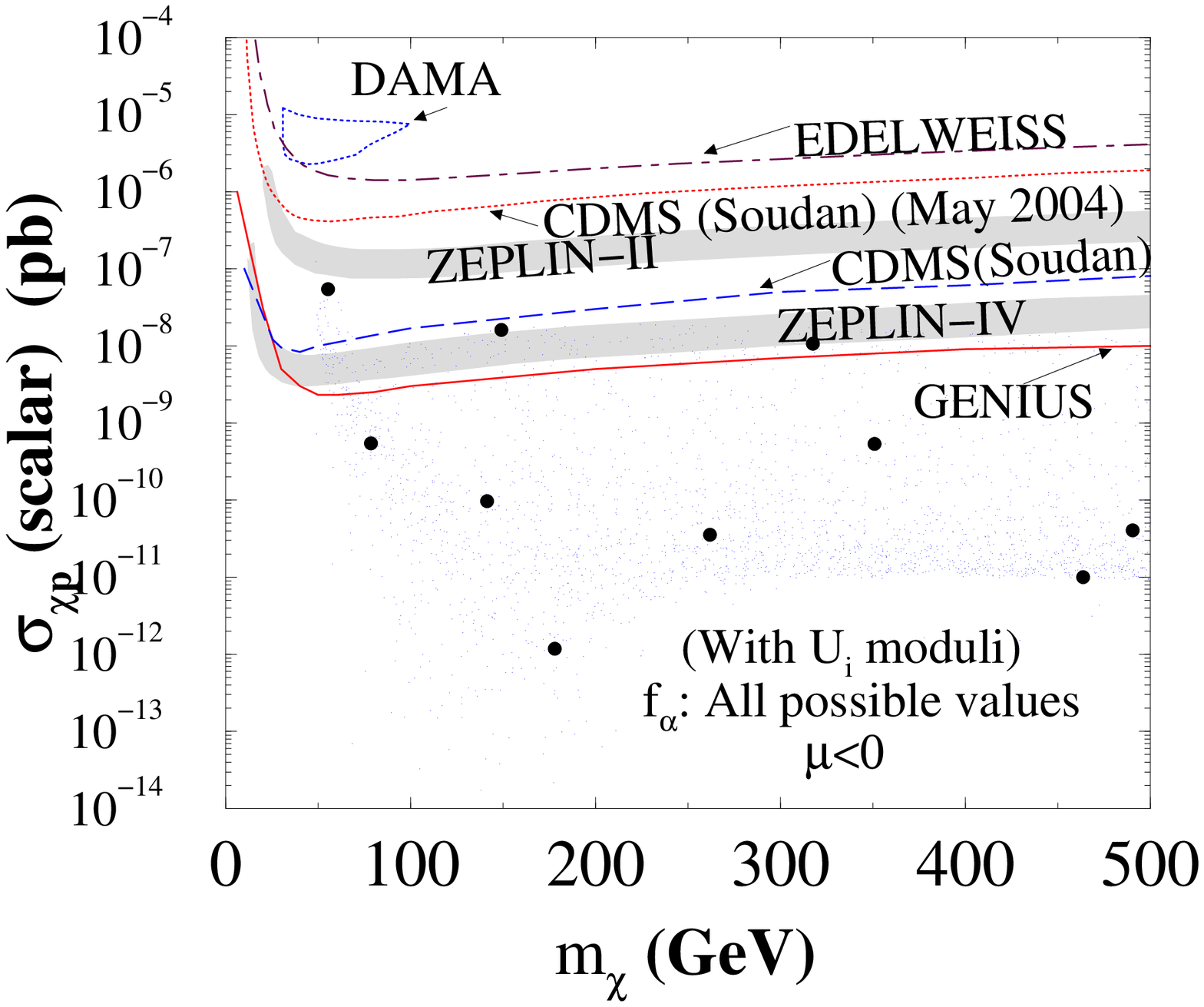} 
\vspace{2mm}
\end{minipage}}
\caption{}    
\label{neg_umod}                   
\end{figure}

 We now analyze the effect of including the complex structure moduli 
$U_i$.  In Fig.(\ref{pos_relic_umod}) we present a composite analysis for 
relic density and $Br(b\rightarrow s+\gamma)$ 
in $(\gamma_s - \mgravi)$ plane for $\mu>0$. 
Here we include $U_i$ moduli and we scan over the self dual points corresponding 
to all possible values of $f_\alpha: 2^n3^{3-n/2},n=0,...6$. This generates 
maximally allowed regions since we are integrating over all allowed  values of
$f_\alpha$. The $b \rightarrow s + \gamma$ contour 
is shown as a dot-dashed line below which the region is maximally  
disallowed. The discarded region by $b \rightarrow s + \gamma$ constraint 
extends up to $\mgravi=750$~GeV.     
The WMAP satisfied relic density region is maximally shown as small 
 shaded area in black and the region is not much different from 
the result of Fig.(\ref{gamma_gravi}) and Fig.(\ref{gamma_gravi_fnew}). 
Similarly, Fig.(\ref{pos_sigmasi_umod}) does not show much different  
spin independent 
LSP-proton cross sections in comparison to Fig.(\ref{sigma_si}) and 
Fig.(\ref{sigma_si_fnew}). We do a similar analysis for relic density 
for $\mu<0$ in Fig.(\ref{neg_relic_umod}). The result shows that 
the WMAP allowed region lies in the range $\gamma_s=0.1-0.2$  which is 
more constraining towards smaller $\gamma_s$ values in comparison 
with the similar cases of Fig.(\ref{min_gamma_gravi}) and 
Fig.(\ref{min_gamma_gravi_fnew}).  The spin independent 
detection cross section result is shown in Fig.(\ref{neg_sigmasi_umod}).
In this case most of the points consistent with WMAP constraint lie at values
of $\mx1>>500$~GeV and fall outside the displayed range. Thus the number
of point consistent with the WMAP constraint lying below $\mx1 =500$~GeV
is very sparce as is evident from Fig.(\ref{neg_sigmasi_umod})
where a few scattered points shown in black circles satisfies the WMAP 
constraint.  However, the range of detection cross sections for 
WMAP satisfying points is not different from those of  
Fig.(\ref{min_sigma_si}) and Fig.(\ref{min_sigma_si_fnew}).  Thus we conclude 
that the inclusion of the complex structure moduli $U_i$ does not 
cause any significant alteration in sparticle mass limits or   
detection cross section than the results without them. 
Another interesting observation is that the case $\mu>0$ shows
that the region consistent with the WMAP  constraint is the
dilaton dominated region where $|\gamma_s|>0.5$ while the case
$\mu<0$ shows that the region consistent with the WMAP constraint
is the moduli dominated region where $|\gamma_s|<0.5$.  Inclusion of 
$U_i$ moduli for $\mu<0$ further restricts $\gamma_s$ so that 
$|\gamma_s|<0.2$.  It should
be interesting to see if this phenomenon is a generic feature in a wider class
of string based models.

The analysis above shows that dark matter
that results in string models with determined $\tan\beta$ is  
much more stringently constrained than in models where
$\tan\beta$ is taken as a free parameter (For a sample of 
recent analyses see 
Refs.\cite{ccn2,Chattopadhyay:2003yk,Baer:2003ru,Binetruy:2003yf,gomez} 
and for a recent review see Ref.\cite{Munoz:2003gx} ). 
While the above analysis is carried out in the framework of heterotic
string models,  similar considerations may apply to other classes of
string models. Thus recently there has been a great deal of activity 
in the intersecting brane models (see, e.g, 
Ref.\cite{Blumenhagen:2001te,Ibanez:2001nd} and the references therein)
and specifically semi realistic models with 3 generations and 
 N=1  supersymmetry have been constructed\cite{Cvetic:2001tj}. 
 The pattern of supersymmetry breaking 
 in a broad class of such intersecting D brane models has
 been analyzed in Ref.~(\cite{Kors:2003wf,Grana:2003ek})  and  some 
common features with the 
soft breaking in heterotic string models are shown to emerge. 
Further, one expects  that
the  constraint of radiative breaking of the electroweak symmetry
in the intersecting D brane models would be very similar to the one in 
the heterotic string models allowing once again a 
determination of $\tan\beta$. Thus we also expect upper limits
on sparticle masses to emerge from the WMAP relic density constraint
for the intersecting D brane models at least for the $\mu$ 
positive case.

\section{Conclusion}
In conclusion, in this paper we have analyzed the implications of  
modular invariant
soft breaking in a generic heterotic string scenario under the constraint
of radiative breaking of the electroweak symmetry. We have emphasized 
the importance of  dilaton and moduli dependent front factors which are
shown to be essential in achieving a modular invariant $V_{soft}$.  
Several forms of the soft parameters $A^0$ and $B^0$ are exhibited,
including the forms which exhibit explicitly their modular weights.
It is shown that in models of this
type $\tan\beta$ is no longer an arbitrary parameter but a 
determined quantity. The constraints of modular invariance along with
a determined $\tan\beta$ define the allowed parameter space very
sharply. Quite interestingly one finds that this parameter space
allows for the satisfaction of the accurate relic  density constraints
given by WMAP. Further, the WMAP constraint combined with the 
FCNC constraint puts upper limits on the sparticle masses for the
case $\mu>0$ which are remarkably low for a class of models 
implying that essentially all of the sparticles would
be accessible at the LHC. Quite remarkably some of the sparticle 
spectrum should also be accessible at the Tevatron. 
Further, an analysis of the neutralino proton cross section
indicates that this cross section should be accessible at the
future dark matter experiments. 
The analysis presented here reveals that for the class of models  
considered
the region of the parameter space consistent with the WMAP constraints
is dilaton dominated for $\mu>0$ and moduli dominated for $\mu<0$.
It should be  interesting to see if this result holds for 
 a broader class of  string based models where $\tan\beta$ 
is again fixed  by the constraints of duality and the radiative
breaking of the electroweak  symmetry.
Finally, we emphasize that the  analysis presented here is the first 
work where the constraint of a determined $\tan\beta$ arising from the
dual constraints of modular invariance and radiative breaking of the 
electroweak symmetry is utilized for the analysis of sparticle masses
and dark matter. The predictions of this analysis on sparticle mass 
upper limits on the size of the neutralino-proton cross section  have
important implications for the discovery of sparticles at the Tevatron,
at the LHC and for  the discovery of dark matter via direct detection.
Further, it would also be interesting to explore the implications
of this predictive modular invariant scenario for other low energy
phenomena, and for the indirect detection of dark matter including possible 
future experiments such as EUSO and OWL\cite{Anchordoqui:2004qh}.\\

\noindent
{\bf Acknowledgements}\\
The authors thank Tomasz Taylor for helpful discussions and thank
 Achille Corsetti for collaboration in the early stages of this work. 
 U.C. is grateful for the hospitality received from the   
Theory Division of CERN where part of this work 
was carried out during his visit.
This work is supported in part by NSF grant PHY-0139967.

\newpage

\newpage


\newpage

\newpage

\newpage

\newpage


\newpage

\newpage

\newpage


\newpage



\begin{thebibliography}{999}

\bibitem{bennett}
C.~L.~Bennett {\it et al.},
Astrophys. J. Suppl. {\bf 148}, 1 (2003), [arXiv:astro-ph/0302207].

\bibitem{spergel}
D.~N.~Spergel {\it et al.},
Astrophys. J. Suppl. {\bf 148}, 175 (2003), [arXiv:astro-ph/0302209].

\bibitem{Chattopadhyay:2003xi}
U.~Chattopadhyay, A.~Corsetti and P.~Nath,
Phys.\ Rev.\ D {\bf 68}, 035005 (2003)
[arXiv:hep-ph/0303201].

\bibitem{elliswmap}
J.~Ellis, K.~A.~Olive, Y.~Santoso and V.~C.~Spanos,
Phys. Lett. {\bf B565}, 176, (2003), [arXiv:hep-ph/0303043];
H.~Baer and C.~Balazs,
JCAP {\bf 0305}, 006 (2003), [arXiv:hep-ph/0303114];
A.~B.~Lahanas and D.~V.~Nanopoulos,
Phys. Lett. {\bf B568}, 55, (2003) [arXiv:hep-ph/0303130].

\bibitem{hb/fp}
H.~Baer, C.~Balazs, A.~Belyaev and J.~O'Farrill,
JCAP {\bf 0309}, 007 (2003), [arXiv:hep-ph/0305191];
H.~Baer, C.~Balazs, A.~Belyaev, T.~Krupovnickas and X.~Tata,
JHEP {\bf 0306}, 054 (2003) [arXiv:hep-ph/0304303];
R.~Arnowitt, B.~Dutta and B.~Hu,
arXiv:hep-ph/0310103;
J.~R.~Ellis, K.~A.~Olive, Y.~Santoso and V.~C.~Spanos,
Phys.\ Lett.\ B {\bf 573}, 162 (2003)
[arXiv:hep-ph/0305212];
J.D. Vergados, P. Quentin and D. Strottman, hep-ph/0310365.

\bibitem{Chattopadhyay:2003qh}
For a review  see,
U.~Chattopadhyay, A.~Corsetti and P.~Nath,
arXiv:hep-ph/0310228;
A.~B.~Lahanas, N.~E.~Mavromatos and D.~V.~Nanopoulos,
Int.\ J.\ Mod.\ Phys.\ D {\bf 12}, 1529 (2003)
[arXiv:hep-ph/0308251].

\bibitem{Chan:1997bi}
K.~L.~Chan, U.~Chattopadhyay and P.~Nath,
Phys.\ Rev.\ D {\bf 58}, 096004 (1998)
[arXiv:hep-ph/9710473];.
J.~L.~Feng, K.~T.~Matchev and T.~Moroi,
Phys.\ Rev.\ D {\bf 61}, 075005 (2000).

\bibitem{nathtaylor}
P.~Nath and T.~R.~Taylor,
Phys.\ Lett.\ B {\bf 548}, 77 (2002)
[arXiv:hep-ph/0209282].

\bibitem{heterotic}
For a sample of heterotic string models see,
L.E.Ibanez, H. P. Nilles and F. Quevedo, Nucl. Phys. {\bf B307}, 109 (1988);
A. Antoniadis, J.Ellis, J. Hagelin, and D.V. Nanopoulos, Phys. Lett.
{\bf B194}, 231(1987); B. R. Green, K. H. Kirklin, P.J. Miron G.G. Ross,
Nucl. Phys.{\bf B292}, 606(1987); R. Arnowitt and P. Nath, Phys. Rev.
{\bf D40}, 191(1989); A.~H.~Chamseddine and M.~Quiros,
Nucl.\  Phys.\  B {\bf 316}, 101(1989);
D.C. Lewellen, Nucl. Phys. {\bf B337}, 61(1990); 
A. Farragi, Phys. Lett. {\bf B278}, 131(1992);
S. Chaudhuri,
S.-W. Chung, G. Hockney, and J.D. Lykken, Nucl. Phys.{\bf
452}, 89(1995); G.B. Cleaver, Nucl. Phys. {\bf B456}, 219(1995); M.
Cvetic and P. Langacker, Phys. Rev. {\bf D54}, 3570(1996); Z.
Kakushadze and S.H.H. Tye, Phys. Rev. {\bf D55}, 7896(1997).

\bibitem{fmtv}
S.~Ferrara, N.~Magnoli, T.~R.~Taylor and G.~Veneziano,
Phys.\ Lett.\ B {\bf 245}, 409(1990);
A.~Font, L.~E.~Ibanez, D.~L\"ust and F.~Quevedo,
Phys.\ Lett.\ B {\bf 245}, 401(1990);
H.~P.~Nilles and M.~Olechowski,
Phys.\ Lett.\ B {\bf 248}, 268(1990);
P.~Binetruy and M.~K.~Gaillard,
Phys.\ Lett.\ B {\bf 253}, 119(1991);
M.~Cvetic, A.~Font, L.~E.~Ibanez, D.~L\"ust and F.~Quevedo,
Nucl.\ Phys.\ B {\bf 361}, 194(1991).

\bibitem{brignole}
A.~Brignole, L.~E.~Ibanez, C.~Munoz and C.~Scheich,
Z.\ Phys.\ C {\bf 74}, 157 (1997);
B.~de Carlos, J.~A.~Casas and C.~Munoz,
Nucl.\ Phys.\ B {\bf 399}, 623(1993);
A.~Brignole, L.~E.~Ibanez and C.~Munoz,
Phys.\ Lett.\ B {\bf 387},769(1996).

\bibitem{nilles}
H.~P.~Nilles,
Phys.\ Lett.\ B {\bf 115}, 193(1982);
S.~Ferrara, L.~Girardello and H.~P.~Nilles,
Phys.\ Lett.\ B {\bf 125}, 457(1983);
M.~Dine, R.~Rohm, N.~Seiberg and E.~Witten,
Phys.\ Lett.\ B {\bf 156}, 55(1985);
C.~Kounnas and M.~Porrati,
Phys.\ Lett.\ B {\bf 191}, 91(1987).

\bibitem{gaillard}
P.~Binetruy, M.~K.~Gaillard and B.~D.~Nelson,
Nucl.\ Phys.\ B {\bf 604}, 32(2001);
M.~K.~Gaillard, B.~D.~Nelson and Y.~Y.~Wu,
Phys.\ Lett.\ B {\bf 459}, 549(1999);
M.~K.~Gaillard and J.~Giedt,
Nucl.\ Phys.\ B {\bf 636}, 365(2002).

\bibitem{kane}
G.~L.~Kane, J.~Lykken, S.~Mrenna, B.~D.~Nelson, L.~T.~Wang and T.~T.~Wang,
Phys.\ Rev.\ D {\bf 67}, 045008 (2003)
[arXiv:hep-ph/0209061]; 
B.~C.~Allanach, S.~F.~King and D.~A.~J.~Rayner,
arXiv:hep-ph/0403255.

\bibitem{can}
A.~H.~Chamseddine, R.~Arnowitt and P.~Nath,
Phys.\ Rev.\ Lett.\  {\bf 49}, 970(1982);
R.~Barbieri, S.~Ferrara and C.~A.~Savoy,
Phys.\ Lett.\ B {\bf 119}, 343(1982).
L. Hall, J. Lykken, and S. Weinberg,
\Journal{\PRD}{27}{2359}{1983}:~ P. Nath, R. Arnowitt and A.H. Chamseddine,
\Journal{\NPB}{227}{121}{1983};
P.~Nath, ``Twenty years of SUGRA,''
arXiv:hep-ph/0307123.

\bibitem{antaylor}
G.~Lopes Cardoso and B.~A.~Ovrut,
Nucl.\ Phys.\ B {\bf 369}, 351(1992);
J.~P.~Derendinger, S.~Ferrara, C.~Kounnas and F.~Zwirner,
Nucl.\ Phys.\ B {\bf 372}, 145(1992);
I.~Antoniadis, E.~Gava, K.~S.~Narain and T.~R.~Taylor,
Nucl.\ Phys.\ B {\bf 407}, 706(1993).

\bibitem{lahanas}
A.~B.~Lahanas and D.~V.~Nanopoulos,
Phys.\ Rept.\  {\bf 145}, 1 (1987).

\bibitem{Chattopadhyay:2002jx}
U.~Chattopadhyay and P.~Nath,
Phys.\ Rev.\ D {\bf 66}, 093001 (2002)
[arXiv:hep-ph/0208012].and references therein.

\bibitem{Bennett:2004pv}
G.~W.~Bennett {\it et al.}  [Muon g-2 Collaboration],
Phys.\ Rev.\ Lett.\  {\bf 92}, 161802 (2004)
[arXiv:hep-ex/0401008].

\bibitem{Antoniadis:1994hg}
I.~Antoniadis, E.~Gava, K.~S.~Narain and T.~R.~Taylor,
Nucl.\ Phys.\ B {\bf 432}, 187 (1994)
[arXiv:hep-th/9405024].

\bibitem{an}
R.~Arnowitt and P.~Nath,
Phys.\ Rev.\ Lett.\  {\bf 69}, 725(1992).

\bibitem{bsgamma}
P. Nath and R. Arnowitt, \Journal{\PLB}{336}{395}{1994};
\Journal{\PRL}{74}{4592}{1995};
F.~Borzumati, M.~Drees and M.~Nojiri, \Journal{\PRD}{51}{341}{1995};
H. Baer, M. Brhlik, D. Castano and  X. Tata, \Journal{\PRD}{58}
{015007}{1998}.

\bibitem{bsgammanew}
H.~Baer, A.~Belyaev, T.~Krupovnickas and A.~Mustafayev,
arXiv:hep-ph/0403214; 
K.~i.~Okumura and L.~Roszkowski, Journal of High Energy Physics,
{\bf 0310}, 024 (2003) [arXiv:hep-ph/0308102];
M. Carena, D. Garcia, U. Nierste, C.E.M. Wagner, Phys. Lett.
{\bf B499} 141 (2001);
G. Degrassi, P. Gambino, G.F. Giudice, JHEP 0012, 009 (2000);
M.~Ciuchini, G.~Degrassi, P.~Gambino and G.~F.~Giudice,
Nucl.\ Phys.\ B {\bf 534}, 3 (1998).

\bibitem{gambino}
P. Gambino and M. Misiak, Nucl. Phys. {\bf B611}, 338 (2001);
 P. Gambino and U. Haisch,  JHEP 0110, 020 (2001);
 A.L. Kagan and M. Neubert, Eur. Phys. J. C7, 5(1999).
A.L. Kagan and M. Neubert, Eur. Phys. J. {\bf C27}, 5(1999).

\bibitem{cleo}
K.~Abe {\it et al.}  [Belle Collaboration],
Phys.\ Lett.\ B {\bf 511}, 151 (2001)
[arXiv:hep-ex/0103042];
S. Chen et.al. (CLEO Collaboration), Phys. Rev. Lett. {\bf 87}, 251807
(2001);
R.~Barate {\it et al.}  [ALEPH Collaboration],
Phys.\ Lett.\ B {\bf 429}, 169 (1998).

\bibitem{cdms}
R. Abusaidi et.al., Phys. Rev. Lett.{\bf84}, 5699(2000),
"Exclusion Limits on WIMP-Nucleon Cross-Section
from the Cryogenic Dark Matter Search", CDMS Collaboration preprint
CWRU-P5-00/UCSB-HEP-00-01 and astro-ph/0002471.

\bibitem{genius}
H.V. Klapdor-Kleingrothaus, et.al., 
"GENIUS, A Supersensitive Germanium Detector System for Rare Events: Proposal", MPI-H-V26-1999,
hep-ph/9910205;
H.~V.~Klapdor-Kleingrothaus,
``Search for dark matter by GENIUS-TF and GENIUS,''
Nucl.\ Phys.\ Proc.\ Suppl.\  {\bf 110}, 58 (2002)
[arXiv:hep-ph/0206250].

\bibitem{cline}
D.~Cline,
``Status of the search for supersymmetric dark matter,''
arXiv:astro-ph/0306124.

\bibitem{smith}
D.~R.~Smith and N.~Weiner,
Nucl.\ Phys.\ Proc.\ Suppl.\  {\bf 124}, 197 (2003)
[arXiv:astro-ph/0208403].

\bibitem{cdmsmay2004}
  [CDMS Collaboration],
``First Results from the Cryogenic Dark Matter Search in the Soudan 
Underground Lab,''
arXiv:astro-ph/0405033.

\bibitem{edelweiss}
G.~Chardin {\it et al.}  [EDELWEISS Collaboration],
Nucl.\ Instrum.\ Meth.\ A {\bf 520}, 101 (2004); 
A.~Benoit {\it et al.},
Phys. Lett. {\bf B545}, 43 (2002), [arXiv:astro-ph/0206271].

\bibitem{damaresult}
R.~Bernabei {\it et al.}  [DAMA Collaboration],
Phys.\ Lett.\ B {\bf 480}, 23 (2000).

\bibitem{ccn2}
U.~Chattopadhyay, A.~Corsetti and P.~Nath,
Phys.\ Rev.\ D {\bf 66}, 035003 (2002)
[arXiv:hep-ph/0201001]; M.~E.~Gomez, G.~Lazarides and C.~Pallis,
Phys.\ Rev.\ D {\bf 61}, 123512 (2000)
[arXiv:hep-ph/9907261].

\bibitem{Chattopadhyay:2003yk}
U.~Chattopadhyay and D.~P.~Roy,
Phys.\ Rev.\ D {\bf 68}, 033010 (2003)
[arXiv:hep-ph/0304108].

\bibitem{Baer:2003ru}
H.~Baer, A.~Belyaev, T.~Krupovnickas and X.~Tata,
JHEP {\bf 0402}, 007 (2004)
[arXiv:hep-ph/0311351].

\bibitem{Binetruy:2003yf}
P.~Binetruy, Y.~Mambrini and E.~Nezri,
arXiv:hep-ph/0312155.

\bibitem{gomez}
M.~E.~Gomez, T.~Ibrahim, P.~Nath and S.~Skadhauge,
arXiv:hep-ph/0404025;
T.~Nihei and M.~Sasagawa,
arXiv:hep-ph/0404100;
M.~Argyrou, A.~B.~Lahanas, D.~V.~Nanopoulos and V.~C.~Spanos,
arXiv:hep-ph/0404286.

\bibitem{Munoz:2003gx}
C.~Munoz,
arXiv:hep-ph/0309346.

\bibitem{Blumenhagen:2001te}
R.~Blumenhagen, B.~K\"ors, D.~L\"ust and T.~Ott,
Nucl.\ Phys.\ B {\bf 616}, 3(2001).
[hep-th/0107138].

\bibitem{Ibanez:2001nd}
L.~E.~Ibanez, F.~Marchesano and R.~Rabadan,
JHEP {\bf 0111}, 002(2001)
[hep-th/0105155].

\bibitem{Cvetic:2001tj}
M.~Cvetic, G.~Shiu and A.~M.~Uranga,
Phys.\ Rev.\ Lett.\  {\bf 87}, 201801(2001)
[hep-th/0107143].

\bibitem{Kors:2003wf}
B.~Kors and P.~Nath,
Nucl.\ Phys.\ B {\bf 681}, 77 (2004)
[arXiv:hep-th/0309167].

\bibitem{Grana:2003ek}
M.~Grana, T.~W.~Grimm, H.~Jockers and J.~Louis,
arXiv:hep-th/0312232.


\bibitem{Anchordoqui:2004qh}
L.~Anchordoqui, H.~Goldberg and P.~Nath,
arXiv:hep-ph/0403115 (To appear in Phys. Rev. D).


\end{thebibliography}
\end{document}